# Structural, Rheological and Dynamic Aspects of Hydrogen-Bonding Molecular Liquids: Aqueous Solutions of Hydrotropic *tert*-Butyl Alcohol

Jure Cerar,[a] Andrej Jamnik,[a] Ildikó Pethes,[b] László Temleitner,[b] László Pusztai [b,c] and Matija Tomšič.[a,*]

[a]Faculty of Chemistry and Chemical Technology, University of Ljubljana, Večna pot 113, SI-1000 Ljubljana, Slovenia.

[b]Wigner Research Centre for Physics, Hungarian Academy of Sciences, Budapest, Konkoly Thege út 29-33., H-1121, Hungary.

[c]International Research Organisation for Advanced Science and Technology (IROAST), Kumamoto University, 2 39 1 Kurokami, Chuo-ku, Kumamoto 860-8555, Japan.

*Correspondence phone and e-mail: +386 1 479 8515 and Matija.Tomsic@fkkt.uni-lj.si

**Abstract**
*Hypothesis:* The structural details, viscosity trends and dynamic phenomena in *t*-butanol/water solutions are closely related on the molecular scales across the entire composition range. Utilizing the experimental small- and wide-angle x-ray scattering (SWAXS) method, molecular dynamics (MD) simulations and the 'complemented-system approach' method developed in our group it is possible to comprehensively describe the structure-viscosity-dynamics relationship in such structurally versatile hydrogen-bonded molecular liquids, as well as in similar, self-assembling systems with pronounced molecular and supramolecular structures at the intra-, inter-, and supra-molecular scales.
*Experiments:* The SWAXS and x-ray diffraction experiments and MD simulations were performed for aqueous *t*-butanol solutions at 25 °C. Literature viscosity and self-diffusion data were also used.
*Findings:* The interpretive power of the proposed scheme was demonstrated by the extensive and diverse results obtained for aqueous *t*-butanol solutions across the whole concentration range. Four composition ranges with qualitatively different structures and viscosity trends were revealed. The experimental and calculated zero-shear viscosities and molecular self-diffusion coefficients were successfully related to the corresponding structural details. The hydrogen bonds that were, along with hydrophobic effects, recognized as the most important driving force for the formation of *t*-butanol aggregates, show intriguing lifetime trends and thermodynamic properties of their formation.

## 1. INTRODUCTION

The process of solvation and the mutual dissolution of hydrophobic and hydrophilic compounds have raised numerous fundamental physico-chemical questions over many years and continue to attract the attention of chemists and physicists [1-5]. The knowledge associated with these topics constitutes an important basis for modern colloidal chemistry, which is typically concerned with interfacial processes, particle or system dynamics and the structure of interphases in the colloidal domain.

The mutual dissolution of hydrophobic and hydrophilic compounds is commonly achieved by introducing amphiphilic molecules into the system. These have both hydrophobic and hydrophilic characters and stabilize the interface between two otherwise immiscible phases. Such a dual molecular nature is sometimes also referred to as the 'Janus effect'[6-8]. The addition of the amphiphilic components usually results in molecular self-assembly, leading to supramolecular structures, *e.g.*, micelles, liposomes, liquid-crystalline phases, etc. The first studies of these and related phenomena reach as far back as the 18$^{th}$ century and are now well understood. Nevertheless, the combination of modern experimental and theoretical approaches offer detailed, new insights into these topics.

Recently, attention has turned to substances with small molecules exhibiting only a weak amphiphilic character that are sometimes addressed as 'amphiphilic solvents' or 'solvo-surfactants' [9, 10]. They can be seen as liquids exhibiting the properties of surfactants, such as surface activity, self-assembly in water, co-micellization, etc. The term 'hydrotrope' is also very commonly used, even though, as reviewed by Kunz *et al.* [11], it comprises a broader range of amphiphilic compounds [12-14]. Their ternary systems are often called 'surfactant-free microemulsions' [15] or 'detergentless microemulsions' [16]. More recently Zemb *et al.* proposed the term 'ultraflexible microemulsions' [17], as such systems exhibit low bending rigidities.

The binary aqueous solutions of such substances are especially interesting due to their strong non-ideal behavior, *e.g.*, in terms of their viscosity, partial molar volume, excess thermodynamic properties, compressibility, and sound-attenuation coefficient [18-23]. Such behavior arises because the hydrophilic part of the molecule forms strong hydrogen-bonds (HBs) with water, while the hydrophobic part introduces a perturbation to the water's hydrogen-bonded structure and induces cooperative ordering, stemming from the hydrophobic hydration effects. However, a full understanding of the hydration structure of such systems and their non-ideal behavior is still incomplete.

In our recent research we have strived to learn more about the structure [24-29] and its effects on the rheological properties [30, 31] and the molecular dynamics in hydrogen-bonding liquids [30-33]. Such knowledge is important for both basic physico-chemical research, often investigating the competition between different types of intermolecular interactions, as well as for applied chemical research, often involving the structure-function relationship and its tuning (*e.g.*, such binary systems are widely used as chemical reaction media, solvents for liquid extraction, chromatography, etc.). Correspondingly, the first objective of this paper is to represent the interpretive power of a comprehensive combination of modern experimental and theoretical techniques to explain the structure-viscosity-dynamics relationship in such systems, *i.e.*, the combination of the experimental small- and wide-angle x-ray scattering (SWAXS) method and the molecular-dynamics (MD) simulations followed by the theoretical calculations of the SWAXS intensities according to the 'complemented-system approach' developed in our group [28]. In addition to this we have x-ray diffraction (XRD), rheological and self-diffusion data and results. The second, no less important, objective of this paper is to reveal and elaborate on the complex structure-viscosity-dynamics relationship in aqueous solutions of 2-methylpropan-2-ol with the trivial name *t*-butanol (TBA) across the whole concentration range in a single study at intra-, inter- and supra-molecular levels of detail. TBA is an important chemical with rich applicability [34-42]. As the presented results reveal, it is also a very interesting structurally and dynamically versatile model system of hydrogen-bonding molecular liquids.

TBA is known to behave as a hydrotrope [43-45]. Aqueous solutions of TBA have the most pronounced non-ideal behavior among the homologous simple alcohols. Further, among the four isomers of butanol (others being: *n*-butanol (butan-1-ol), *sec*-butanol (butan-2-ol) and iso-butanol (2-metylpropan-1-ol)), it is the only isomer that is completely miscible in water under ambient conditions. Numerous available studies of TBA systems indicate that an investigation of the structure of aqueous TBA solutions is far from being a trivial task. The majority of them are performed in a rather limited (although some are broader) concentration range [44, 46-48], but we could not find a structural study of aqueous TBA that would really closely combine the x-ray scattering experiment and theory across the whole concentration range. Both experimental [47, 49] and simulation studies [6, 7, 46, 50-58] show that mixtures of TBA and water are homogeneous on the mesoscopic scale [44, 59], while on the microscopic scale heterogeneities can be observed – a feature strongly dependent on the composition of the solution. Such heterogeneities are associated with the formation of discrete local TBA-rich and water-rich regions, whose effective sizes and structure remain unclear. Some even claimed the presence of meso-scale inhomogeneities (200 nm and more), which were later shown to originate in impurities [44, 47, 60].

MD computer simulations based on a reliable molecular model enable a direct insight into the system and represent an important tool for detailed structural studies on the molecular and supra-molecular levels. In practice, the simplified molecular model and the limitations of the available computing power (limited number of simulated molecules) can bring ambiguities to the results of various studies, *e.g.*, Gupta and Patey [50] showed that some force-field models of aqueous TBA suggest a phase separation, which is clearly inconsistent with the experimental findings, while others show the presence of smaller TBA-rich aggregates that are either stable or slowly coalesce. Furthermore, in small simulation boxes some models suggested the formation of molecular TBA aggregates, but when a larger-scale simulation was performed (64,000 particles) the same models led to macroscopic phase separation. This posts two crucial questions regarding the simulations of aqueous TBA: (*i*) is the simulation box big enough to correctly consider the long-range intermolecular interactions and (*ii*) which force field is most appropriate for this system [50, 51, 57]?

In this study we ensured credible results for the structure of aqueous TBA by a successful comparison of the theoretical scattering functions from the simulation [24, 28] and the experimental SWAXS data. The complemented-system approach allows us to critically assess the accuracy of the force-field model to reproduce the intermolecular and supramolecular structures of the real system [32, 33] in the whole concentration range. Comparing the total structure factors from the MD-simulation results and the reduced experimental x-ray diffraction (XRD) data a validation of the model was also made at the intramolecular structural level. Furthermore, we performed MD simulations in large-scale simulation boxes (100,000 and even 3,000,000 particles) in order to eliminate the finite-box-size effects. Furthermore, we offer a new insight into the dynamic properties of aqueous TBA systems through an analysis of the HB lifetime and the calculated standard thermodynamic properties of HB formation. We also explored the viscous properties of the aqueous TBA system and their relation to its supra-molecular structure.

## 2. EXPERIMENTAL AND METHODS

The details of the chemicals, the SWAXS [61, 62] and XRD [63, 64] measurements, the molecular-dynamics simulations [65-82] for the study of the structure and dynamics (GROMACS software package version 2018.2 [83-85]; OPLS-AA *tert*-butyl model [86-88] and TIP4P or TIP4P/2005 water models [89, 90]), the statistical analysis of the BHs [91] for their lifetimes [92] and the standard thermodynamic properties [93], and the non-equilibrium molecular-dynamics periodic perturbation method for the viscosity study [94-96] based on the Carreau-Yasuda model extrapolation [96] are given in the online Supporting Information (SI). In the following, we only briefly describe the SWAXS-intensity-calculation procedure according to the 'complemented-system approach'. Throughout the paper, the composition of the samples is presented as a molar fraction of the TBA, $x_{\text{TBA}}$.

**The complemented-system approach.**

The theoretical SWAXS intensities were calculated from the MD-simulation results using the complemented-system approach [28]. It is based on the well-known Debye equation, extended with an additional term representing the interference of the limited simulated system with the infinite surroundings of an electron density equal to the averaged electron density of the simulated system [28]:

$$\frac{d\Sigma}{d\Omega}(q) = \frac{1}{V} \left\{ \sum_{i=1}^{N}\sum_{j=1}^{N} b_i(q) b_j(q) \frac{\sin(qr_{ij})}{qr_{ij}} H(r_c - r_{ij}) \right\} + \left( \sum_{k=1}^{N_T} b_k(q) \rho_k \right)^2 \left[ (2\pi)^3 \delta(q) - \frac{4\pi}{q^3}(\sin(qr_c) - qr_c \cos(qr_c)) \right],$$

(1)

where $d\Sigma/d\Omega$ denotes the differential scattering cross-section per unit volume, $V$ is the volume of the simulation box, $N$ is the number of atoms present in the system, $r_{ij}$ is the distance between the particles $i$ and $j$, $b_i$, $b_j$, and $b_k$ are the corresponding atom-type structure factors [97], $r_c$ is the cut-off distance (half of the simulation-box length), $H$ is the Heaviside step function, $N_T$ is the total number of particle types, $\rho_k$ is the density of the corresponding pseudo-atom type and $\delta$ is the Dirac delta function. The different partial contributions to the total scattering intensities (theoretical analogy to 'contrast matching') can also be calculated utilizing eqn (1) and the simulation boxes with atoms of certain types being removed – more details are given in the SI. In order to be able to directly compare the theoretical and experimental SWAXS intensities, the former need to be numerically smeared [24, 28]. For a detailed description of the complemented-system approach the reader is referred to our original article [28] and to other studies that demonstrate the successful application of this method [29, 30, 32, 33].

## 3. RESULTS AND DISCUSSION

To keep the discussion at a general level, we will only present the main findings of this study below and in parallel provide some more detailed explanations in the SI.

### 3.1. Structural aspects of the TBA/Water System

As the experimental SWAXS and XRD data embody realistic structural information about the studied system, they represent the very basis of our research. In Fig. 1a the experimental and calculated SWAXS intensities for the TBA/water mixtures with molar fractions of TBA, $x_{TBA}$, in the whole concentration range ($0 \leq x_{TBA} \leq 1$) are compared. The experimental uncertainty of the data is well within the symbol size. In the experimental SWAXS curve of the pure TBA two distinct scattering peaks can be seen, *i.e.*, the peak at $q \sim 7.4$ nm$^{-1}$ and the peak at $q \sim 13.1$ nm$^{-1}$.

Throughout the following text we will address them as the '*inner*' and '*outer*' scattering peaks, respectively.

These two peaks are characteristic for primary as well as other alcohols and were already discussed in detail in our previous studies [24, 32, 33]. In general, the inner scattering peak corresponds to the correlations between the hydroxyl groups of the neighboring –OH skeletons [24, 98], which represent the hydrophilic regions in the system, whereas the outer scattering peak corresponds to the correlation lengths within the hydrophobic regions. These are composed of the alkyl tails occupying the space between the sequentially hydrogen- bonded –OH groups forming the backbone of the alcohol aggregates (–OH skeletons).

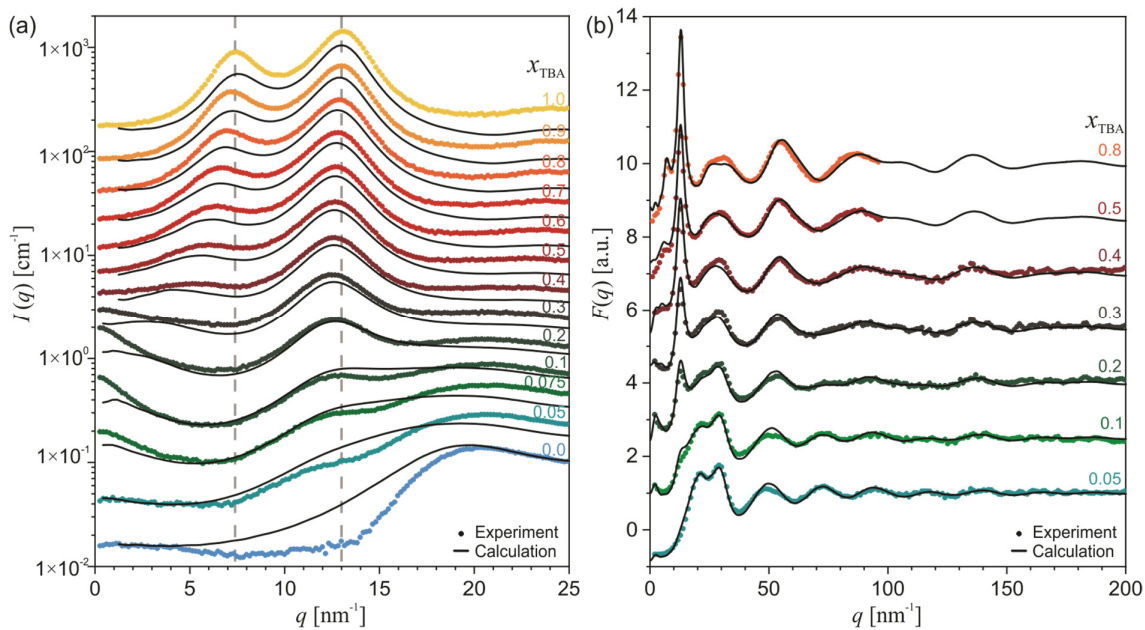

Fig. 1 (a) The experimental and the calculated SWAXS intensities of the TBA/water system across the whole concentration range. The vertical lines are drawn at 7.4 and 13 nm$^{-1}$. (b) The total structure factor, $F(q)$, as obtained from the experiment and as calculated from the model MD simulation results. The data is shifted upwards by (a) multiples of 2 and (b) increments of 1.5 a.u..

By increasing the water concentration in the TBA/water system, the structural changes are evident from the shifts in the scattering-peak positions in Fig. 1a. The inner scattering peak gradually shifts towards lower $q$-values (from 7.43 nm$^{-1}$ at $x_{TBA} = 1$ to 5.35 nm$^{-1}$ at $x_{TBA} = 0.4$ corresponding to real space distances of 0.85 and 1.18 nm, respectively) until it eventually vanishes for the compositions with $x_{TBA} \lesssim 0.4$. Further on, a pronounced, very low-$q$ scattering intensity increase starts to rise and fully develops towards the composition $x_{TBA} \sim 0.2$, but with a further decrease of $x_{TBA}$ it fully disappears at $x_{TBA} \sim 0.05$. In parallel, the outer scattering-peak position gradually shifts from 13.07 at $x_{TBA} = 1$ to 12.51 nm$^{-1}$ at $x_{TBA} = 0.3$ and then interestingly shifts back towards the higher $q$-region to $\sim$12.80 nm$^{-1}$ at $x_{TBA} = 0.05$, which corresponds to the real space distances of 0.48, 0.50, and 0.49 nm, respectively. The intensive increase of the scattering curves in the very low-$q$ region observed in the range $0.35 \gtrsim x_{TBA} \gtrsim 0.05$ indicates significant microscale structuring. Interestingly, these raw SWAXS curves already indicate the existence of a couple of different structural regimes in the TBA/water mixtures, but to obtain more detailed structural information we must turn to the MD-simulation results.

Fig. 1b shows the experimental XRD total structure factors together with the calculated functions obtained on the basis of the MD simulation model's results. We need to remember the reciprocal relationship between real and reciprocal space, *i.e.*, the information regarding the big dimensions are mostly expressed at smaller values of the scattering vector $q$ and vice versa. Therefore, a very good agreement between the calculated and experimental functions in the broad range of the scattering vector in Fig. 1b confirms the credible structure of the modeled TBA and the water molecules on the intramolecular scale and validates the model on this level. Similarly, we can claim that in Fig. 1a, where the SWAXS data also present the range of very low $q$-values in more details, a good qualitative agreement between the calculated and experimental data is observed. For a more detailed explanation the interested reader is directed to the SI, where an additional investigation of the scattering-intensity increase in the very low-$q$ range at moderate TBA concentrations is discussed and shown in Fig. S2 and S3. Eventually, we can conclude that the applied force-field models also successfully reproduce the general structural characteristics of the studied TBA/water system on the inter- and supra-molecular levels and enable us to draw reasonable conclusions about the structural details and the transitions in aqueous TBA.

In this manner the simulated analogy with the contrast-matching experiment [24] was performed and revealed three partial contributions to the overall SWAXS intensities that are shown in Fig. 2. The $I_{\text{TBA}}(q)$ and $I_{\text{H}_2\text{O}}(q)$ represent the scattering of just the TBA and the water molecules, respectively, and $H_{\text{TBA-H}_2\text{O}}(q)$, just their cross-term contribution. The individual sum of the three contributions equals the overall SWAXS intensity in Fig. 1a. The calculations of the partial contributions are explained in details in the SI, where the individual partial contributions to the $I_{\text{TBA}}(q)$ are also presented in Fig. S5.

These partial contributions reveal at least three very important and even surprising structural details of the TBA/water system that we need to address in the following. Firstly, from the $I_{\text{TBA}}(q)$ in Fig. 2a we can see that the inner scattering peak corresponding to the correlations between the neighboring –OH skeletons holds its position while changing the composition of the TBA/water mixture (see also Fig. S5b); a feature that seems to oppose the observed trend of the inner-peak movement in the overall SWAXS curves in Fig. 1. This clearly indicates that the interpretation of the positions of the peaks in the scattering curve of multi-component systems is a very complex task. The intensity of this peak decreases with the increasing water concentration until it disappears for the compositions $x_{\text{TBA}} \lesssim 0.8$. Its constant position indicates rather similar effective correlations between the –OH groups in the backbones of the TBA aggregates in this concentration range. In parallel, the outer scattering peak corresponding to the correlations between the alkyl molecular parts within the local hydrophobic regions in the structure somewhat surprisingly broadens and shifts towards lower $q$-values, *i.e.*, from 13.1 to 11.3 nm$^{-1}$, reflecting the distances of 0.48 nm and 0.56 nm, respectively.

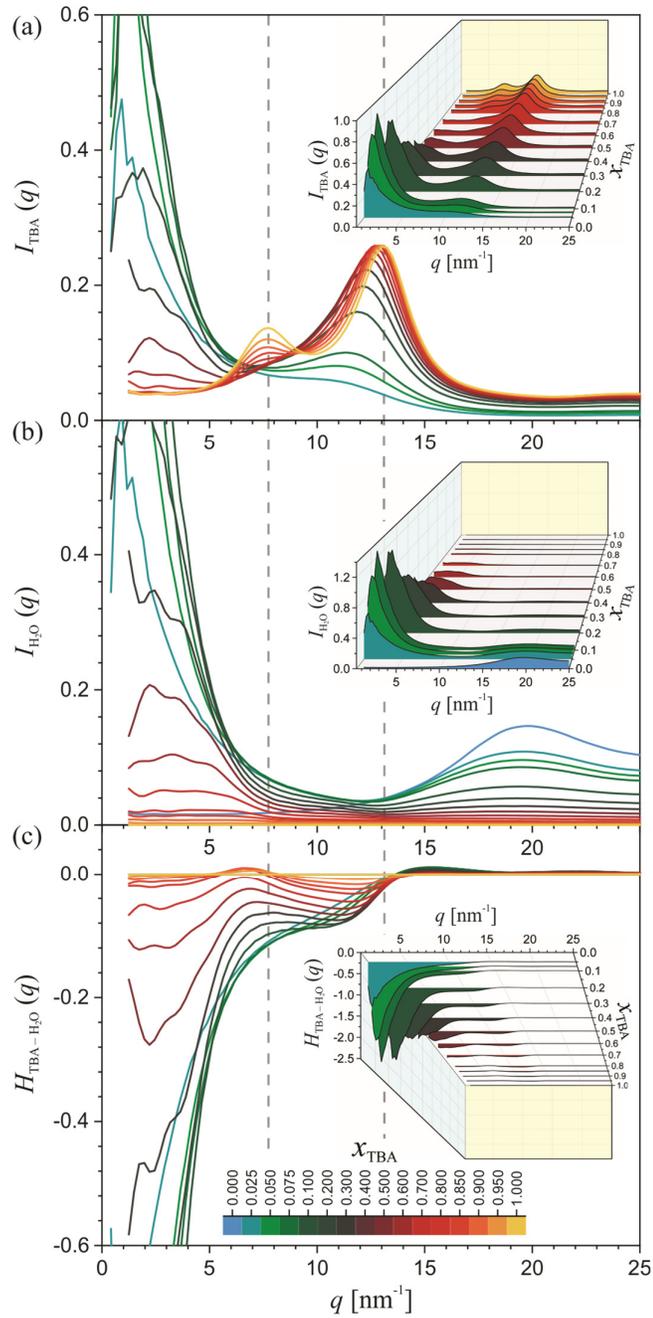

Fig. 2 The partial scattering contributions to the theoretical SWAXS intensities of the TBA/water system at various compositions: (a) $I_{\mathrm{TBA}}(q)$, (b) $I_{\mathrm{H_2O}}(q)$, and (c) $H_{\mathrm{TBA\text{-}H_2O}}(q)$. The vertical lines are drawn at the same positions as in Fig. 1.

The latter is the second very important structural feature in Fig. 2a, which indicates a qualitative change in the effective correlations between the TBA alkyl tails. It is further argued by a very similar peak-position shift of the detailed partial contribution $I_{-C(CH_3)_3}(q)$ presented in Fig. S5a in the SI. It can be reasoned with the effectively increasingly hydrophilic nature of the TBA/water system with increasing water concentration, which obviously profoundly affects the intermolecular packing of the hydrophobic molecular tails. Interestingly, with the increasing hydrophilicity of the system these effective correlations between the alkyl parts of the TBA molecules become somewhat longer-ranged, even though no major changes in the TBA molecular conformations can be inferred from Fig. S6 in the SI.

To explain this highly intriguing phenomenon we further inspect the radial pair-distribution functions $g(r)$ presented in Fig. 3, which reveal the intermolecular correlations in real space. The answer lies in the opposing trend of changes of the first two peaks marked as (*i*) and (*ii*) in Fig. 3a. As it turns out, the first peak corresponds to the closest intermolecular tertiary C–C correlations, resulting from neighboring mutually hydrogen-bonded TBA molecules and the second peak to somewhat more distant tertiary C–C correlations emanating from the neighboring TBA molecules, which are mutually non-hydrogen-bonded and have mutually facially oriented alkyl tails. The spatial distributions (SDs) that lead to those two peaks are presented at the bottom of Fig. 3 and also in a 3D version in three short video clips, Videos S1, S2 and S3, available in the SI. With an increasing water concentration the direct TBA-TBA HB becomes relatively less probable, which is reflected in the reduction of the first peak (*i*). In parallel, due to the increasing hydrophobic effect the direct correlations between the TBA alkyl tails become relatively more probable and the peak (*ii*) gains in height. In turn, this means that the correlations between the TBA alkyl tails occur on effectively larger distances with increasing water concentration – a second (more distant) feature that is reflected in the outer scattering peak in Fig. 2a.

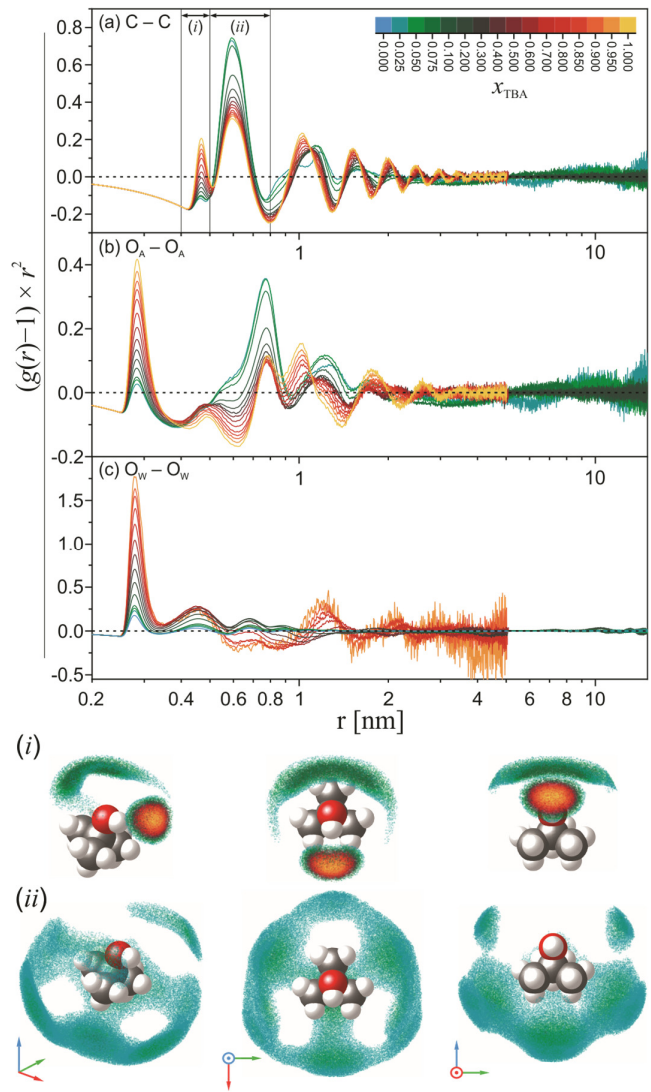

Fig. 3 The radial pair-distribution functions $g(r)$ presented as $(g(r) - 1)) \cdot r^2$ functions for (a) TBA tertiary C–C, (b) TBA O–O and (c) water O–O atoms at various compositions of the TBA/water system. Bottom: The SDs with a double cut-off distance of (*i*) 0.4-0.5 nm and (*ii*) 0.5-0.8 nm, which correspond to the first and second peaks from Fig. 3a, respectively.

The third important structural feature seen in Fig. 2a is connected to the fact that with increasing water concentration, already at the compositions $x_{TBA} \lesssim 0.85$ the $I_{TBA}(q)$ partial scattering contribution gradually develops a sharp intensity increase at very low $q$-values. Interestingly, such an increase cannot be resolved from the total scattering curve in Fig. 1 at such high $x_{TBA}$ at all; it can only be noticed at the compositions $0.35 \gtrsim x_{TBA} \gtrsim 0.05$. This indicates the well-established supramolecular structure of the system also at higher TBA fractions – a feature also evident from the $I_{H_2O}(q)$ in Fig. 2b. However, due to the different sign of the scattering contrast of the two distinct TBA-rich and water-rich local regions, the cross-term contribution $H_{TBA-H_2O}(q)$ in Fig. 2c is strongly negative in the very low-$q$-regime. Obviously, it counterparts the $I_{TBA}(q)$ and $I_{H_2O}(q)$ contributions in this $q$-regime and practically masks such an intensity increase in the overall SWAXS curves at TBA fractions $0.85 \gtrsim x_{TBA} \gtrsim 0.35$, which is not the case below this concentration range any more. Canceling the different scattering contributions has already been observed in some theoretical studies of various alcohol-based systems [52, 99-101]. The results in Fig. 2 show that the overall scattering curves in Fig. 1 result from the intricate interplay of many different contributions – the interested reader is directed to the SI, where we explain these important issues in detail, considering the scattering contrast in the specific TBA/water system (see Fig. S7 in SI) and the well-known Babinet principle [102-104].

To enlighten the structural differences for different system compositions on the supramolecular level, we turn our attention to the schematic representations of the 2-nm-thick slices through the MD-simulation box configurations in Fig. 4a-d, which follow in increasing water concentration in the system. As is evident in Fig. 4a, the water molecules appear to distribute relatively evenly over the system and reside in very close vicinity to the TBA's hydroxyl skeletons in the concentration range $x_{TBA} \gtrsim 0.85$ (the mass fraction of water is less than ~0,04). Something very similar has already been observed in the alcohol-rich binary systems of primary alcohols and was interpreted as the 'swelling' of the –OH skeletons [105]. As it turns out, such water molecules insert themselves between two –OH groups in the backbone of the TBA aggregate, form HBs with them and in this way 'dynamically split' the TBA-aggregate.

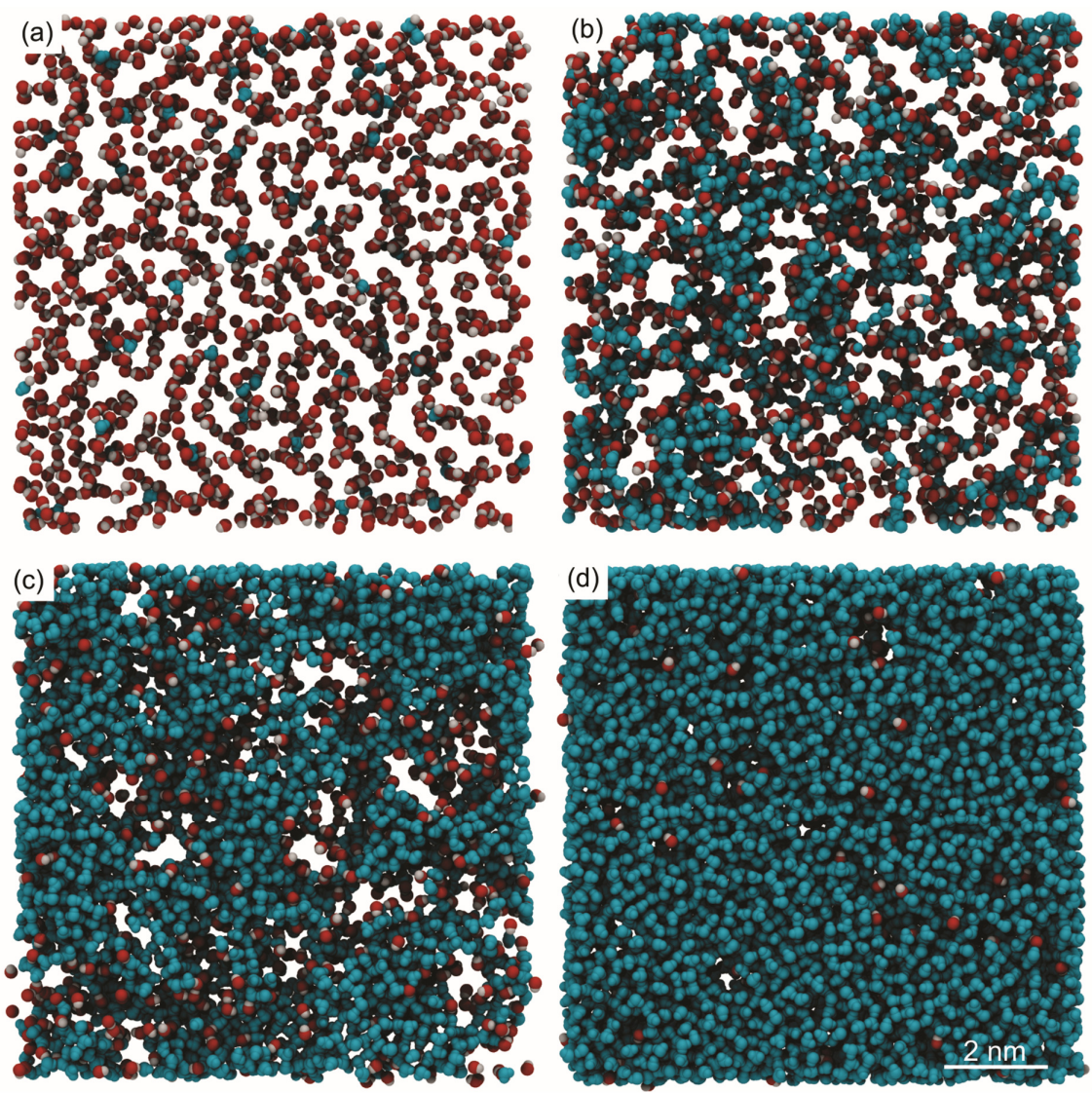

Fig. 4 Schematic visualization of a 2-nm-thick slice through the MD-simulation box for the TBA/water mixture with $x_{TBA}$ of (a) 0.9, (b) 0.5, (c) 0.2 and (d) 0.025. The red beads depict the TBA hydroxyl groups and blue beads the water molecules. The TBA's alkyl groups are omitted. Therefore, the white regions and the voids represent the local hydrophobic regions within the system.

The second, rather broad, concentration range with $0.85 \gtrsim x_{\text{TBA}} \gtrsim 0.35$, where the transition from the TBA-dominated structure to the water-dominated structure occurs over a bicontinuous-like type of the micro-phase-separated structure, is represented in Fig. 4b. In some studies, this range is even referred to as the 'flexible bicontinuous microemulsion' [14, 106].

The third range with $0.35 \gtrsim x_{\text{TBA}} \gtrsim 0.05$, where the distinct, micro-phase-separated TBA-rich local regions within the continuous aqueous phase gradually decrease in size, is presented with the scheme in Fig. 4c. We must mention that at $x_{\text{TBA}} \approx 0.2$ the weight fraction of TBA in the system is already 0.5 and the local TBA-rich regions already reach an almost comparable effective size and volume fraction to the local water-rich regions. This range eventually transits to the fourth structurally characteristic concentration range with $x_{\text{TBA}} \lesssim 0.05$, where mostly the single hydrated TBA molecules are found within the continuous aqueous medium, as depicted in Fig. 4d.

### 3.2. The Rheological Aspects of the TBA/Water System

Comparing the observed transition compositions between the structural regimes and the experimental trends of other available physico-chemical data for the TBA/water system [107-111], we observe a good level of conformance. As the structure and rheological behavior of the system are expected to be closely related, we were surprised not to find a similar comparison already available for the TBA/water system. In the following, we parallel and relate structural results with the observed rheological trends in terms of viscosity, HB characteristics and molecular diffusion coefficients.

On a molecular level, the viscosity represents a measure of the friction/resistance when traversing through the liquid system. During this process the energy is dissipated not only to overcome the intermolecular interactions, but also the steric hindrances, *e.g.*, to reorganize or break up the molecular aggregates or supramolecular structure in the liquid. Therefore, it is not surprising to see that a structured molecular liquid such as the TBA/water system shows a very interesting viscosity behavior. Comparing pure TBA to other pure butanol isomers, it is clear that the viscosity grows considerably with the bulkiness (branching) of the alkyl chain ($\eta_{n-\text{butanol}} = 2.54$ mPa·s, $\eta_{sec-\text{butanol}} = 3.10$ mPa·s, and $\eta_{iso-\text{butanol}} = 3.95$ mPa·s) [112] where the pure TBA system has the highest viscosity ($\eta_{\text{TBA}} = 4.320$ mPa·s) [68]. The size distributions of the molecular aggregates in the TBA/water system are presented in Fig. S8 in the SI and show a decreasing trend with an increasing concentration of water in the system. Interestingly, in *n*-butanol, much larger hydrogen-bonded molecular aggregates were found [24, 32]. This indicates that the bulkiness and rigidity of the alkyl tails have a much stronger effect on the viscosity, as does the size of the aggregates. Nevertheless, the latter plays an important role in the viscosity within a certain system.

In Fig. 5a we present the calculated dynamic zero-shear viscosities for the TBA/water system and compare them to the available experimental data [68]. It has been shown that the TIP4P/2005 water force-field model reproduces the dynamic viscosity of pure water correctly [113]. On the other hand, to the best of our knowledge, the OPLS-AA force field [87] was not optimized to reproduce the viscosity values and clearly underestimates them considerably. Nevertheless, the combination of the two obviously still reproduces the qualitative trend, *i.e.*, the appearance of the maximum and the shallow minimum of the experimental viscosity data (red circles). A surprisingly good qualitative trend with the correct maximum and minimum positions was obtained throughout the whole composition range also with the older version of the water model (TIP4P; blue circles), as both of the models in the combination quantitatively underestimate the viscosity. Nevertheless, we were able to use these models to explain the structure-viscosity relationship in the TBA/water system successfully.

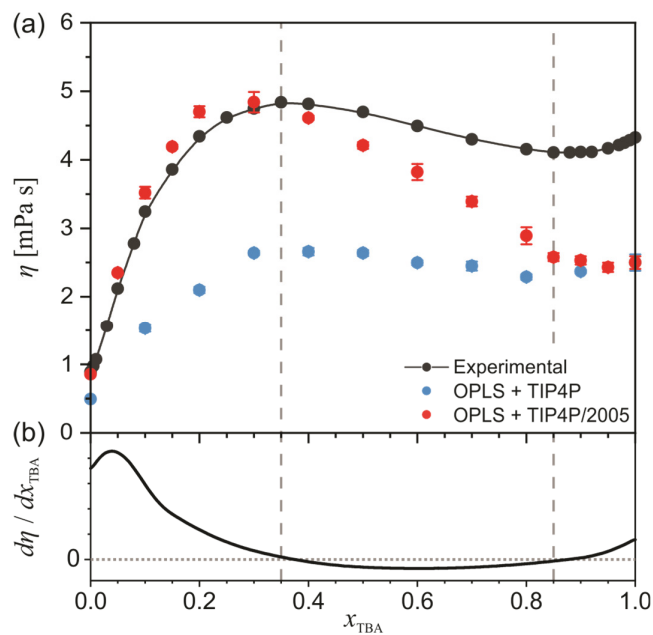

Fig. 5 (a) The experimental and calculated dynamic viscosity values of TBA/water system at different compositions with OPLS-AA+TIP4P and OPLS-AA+TIP4P/2005 force fields. The experimental values were taken from ref. [68]. (b) The first derivative of the cubic B-spline fit to the experimental data.

As is evident in Fig. 5, adding water to the TBA in a low concentration initially results in a slight viscosity decrease that reaches a shallow minimum at $x_{TBA} \sim 0.85$ and with further additions turns to an increasing viscosity trend, reaching the maximum value at $x_{TBA} \sim 0.35$ ($\eta = 4.838$ mPa·s), which even exceeds the viscosity of pure TBA. Further dilution of the TBA shows a steep viscosity decrease towards the viscosity of pure water ($\eta = 0.8902$ mPa·s) [68]. The black curve in Fig. 5b represents the first derivative of the fit to the experimental data and reveals an inflection point of the fit at around $x_{TBA} \sim 0.05$. Interestingly, all these characteristic $x_{TBA}$ values perfectly agree with the transition compositions of the four distinct structural ranges discussed. This proves that the viscosity changes in the TBA/water system are indeed strongly related to the molecular and supramolecular structural changes in the system.

As the molecular aggregates in the TBA/water system, which are expected to influence the viscosity behavior, are primarily governed by the hydrogen-bonding between the TBA molecules, we have further analyzed the HBs in the system. This was based on the model results counting only the 'donor HBs' (maximum one per TBA and 2 per water molecule). To facilitate the interpretation, we show these results in three representations: as the absolute number of HBs in the simulation box of side length 10 nm, $N_{HB}$, in Fig. 6a, as the relative fraction of HBs in the system, $x_{HB}$, in Fig. 6b, and as the average number of HBs per molecule in Fig. 6c. We recognized three different HB types, *i.e.*, TBA-TBA, TBA-water, and water-water HB. For the interested reader, in Fig. S9 in the SI we also provide the results for the average number of HBs per molecule considering both the donor and acceptor HBs.

Fig. 6 clearly reveals a few interesting hydrogen-bonding features: (*i*) the total number of HBs in the system is much lower in the TBA-rich system in comparison to the water-rich system, (*ii*) the number of mixed TBA-water HBs is a maximum at $x_{TBA} \sim 0.35$, (*iii*) but in a relative sense the maximum of the mixed TBA-water HBs is in the middle of the range $0.85 \gtrsim x_{TBA} \gtrsim 0.35$, where the system has a bicontinuous type of structure, (*iv*) up to $x_{TBA} \sim 0.35$ the fraction of the TBA-TBA type of HB is interestingly very low, (*v*) with the addition of water into the system the fraction of water-water type HB adapts the linear trend for $x_{TBA} \lesssim 0.85$, and (*vi*) according to the trends of the $\langle N_{HB}(\text{water})\rangle$ and $\langle N_{HB}(\text{TBA})\rangle$ values, we can argue that obviously the presence of TBA in water promotes the donation of the HBs of water molecules, and vice versa, that the presence of water in the TBA demotes the donation of the HBs of the TBA molecules.

The latter claim conforms to the increasing role of the hydrophobic effect in structuring the TBA molecules with the increasing water concentration inferred from the SWAXS and MD data presented in the previous subchapter. Similarly, at $x_{TBA} \sim 0.3$ the average number of donated

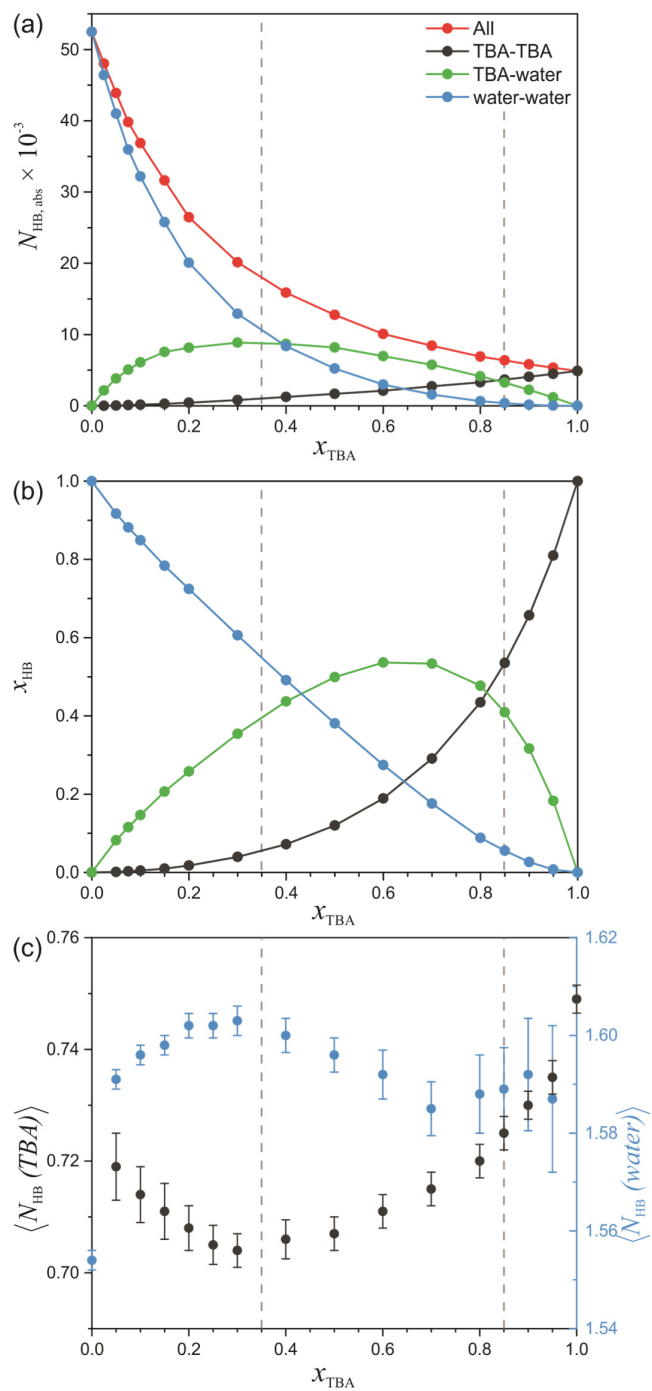

Fig. 6 Composition dependence of the (a) total number of HBs, $N_{HB}$, for different HB types: TBA–TBA, TBA–water, water–water, and all types together, (b) HB fraction, $x_{HB}$, of a certain type and (c) average number of donated HBs per molecule of type X, $\langle N_{HB}(X)\rangle$. Vertical lines denote the compositions $x_{TBA}$ = 0.35 and 0.85.

HBs per molecule in Fig. 6c shows a maximum for water and a minimum for TBA molecules, which is already in the concentration range with discrete local hydrophobic TBA regions. As can also be seen in Fig. 4c, in this regime the TBA molecules orient with the –OH groups towards the interface with water and with the alkyl tails mutually towards each other due to a strong hydrophobic effect that obviously causes the lowest average number of HBs per TBA molecule. However, the increasing trend in $\langle N_{HB}(TBA)\rangle$ with the increasing water concentration on the left-hand side of the minimum in Fig. 6c, is in agreement with the approaching towards an aqueous system with single hydrated TBA molecules – as the local hydrophobic TBA environments reduce in size, the TBA molecules at the interface obviously increase their contact with water, enabling them to donate a somewhat larger fraction of HBs per TBA molecule again. All these facts also prove a very strong relationship between the hydrogen-bonding phenomena and the structural and rheological properties of the TBA/water system.

To check the viscosity effects on the dynamics in the TBA/water system, we calculated the self-diffusion coefficients of the model TBA and the water molecules at 25 °C that are shown in Fig. 7, together with the available Fourier transform pulse gradient spin-echo NMR experimental data for 28 °C [79]. Due to the slight difference in temperatures for these two data sets and the use of deuterated chemicals ($D_2O$ and TBA-$d_{10}$) in the experiments, these data sets cannot be compared quantitatively. At the same time, due to the situation with the reproduction of the viscosity values in Fig. 5a, which are according to the Stokes-Einstein relation [114] in a reciprocal relationship with the diffusion coefficient, we would expect that the TIP4P/2005 model will reproduce the self-diffusion coefficient of water molecules well, but the other models used would have more problems – this is actually clearly seen in Fig. 7. These calculated self-diffusion data have also been used to assess the system's viscosity values, as discussed in the SI and shown in Fig. S10.

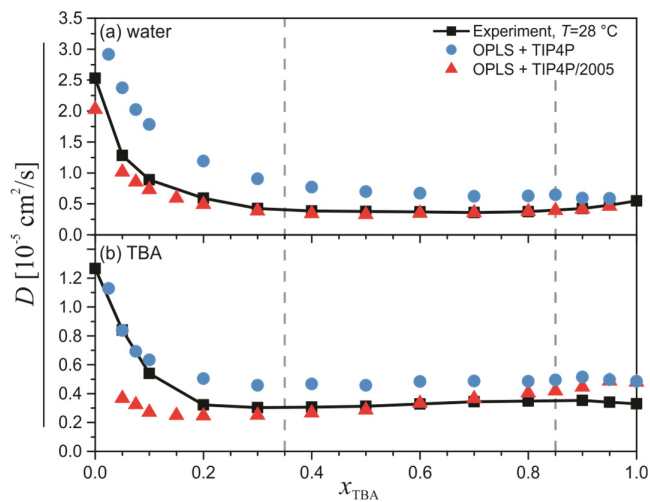

Fig. 7 The experimental and calculated self-diffusion coefficient for (a) water and (b) TBA molecules for different aqueous TBA-system compositions for different models of the molecules. The experimental values were taken from ref. [79].

In any case, the model system is also able to reproduce the general trend of the self-diffusion coefficients reasonably well. The only missing feature is the slight increase in the trend of the calculated water self-diffusion coefficient for the TIP4P model at concentrations $x_{TBA} \gtrsim 0.85$ observed in the experimental data. The calculated $D_{H2O}$ values even seem to decrease slightly there. However, in general we can conclude that the vertical dashed lines in Fig. 7, *i.e.*, the typical transition $x_{TBA}$ values of different structural regimes in the TBA/water system, also conform to the recognized trends of the self-diffusion coefficients.

**3.3. Structure-Viscosity-Dynamics Relationship in the Aqueous TBA System**

Trying to parallel and explain the observed structural, rheological and self-diffusion trends, we can state that with the addition of water to pure TBA in a small concentration ($x_{TBA} \gtrsim 0.85$) the water molecules mostly insert themselves between two –OH groups in the backbone of the TBA aggregates and HB to them. As the small water molecules are correspondingly more diffusive, they dynamically split the TBA aggregates and reduce their size. This in turn results in an initial decrease of the system's viscosity and explains an increase of the TBA self-diffusion coefficient in this range – even though the calculated data in part do not confirm the trend in this range, the experimental data clearly indicate the initial reduction of the water self-diffusion coefficients with the dilution of the system.

This can be reasoned from the fact that such water molecules that split the TBA aggregates form a HB with the latter and in this way somewhat restrict themselves. Indeed, the fraction of the TBA-water type HBs steeply increases in this range and gradually levels-off just around the composition at which the water self-diffusion coefficient turns in trend (see Fig. 6 b and Fig. 7). The fact that the fraction of the water-water type of HBs (representing water molecules that are expectedly more diffusive) in parallel with the dilution of the TBA initially remains relatively low, but starts to increase more steeply just around the composition where the water self-diffusion coefficient trend turns, conforms to this reasoning. Interestingly, just in this concentration range the viscosity of the system turns to an increasing trend with the dilution of the system by water. This is the consequence of the gradual changes in the relative nature of the TBA aggregates – the hydrophobic effect progressively relatively gains in importance, becoming the structural driving force, which happens on account of the decreasing relative importance of the TBA-TBA-type hydrogen bonding with the dilution of the aqueous TBA system. The inflection point of the increasing viscosity trend is around $x_{TBA} \sim 0.6$, where also the maximum in the trend of TBA-water HB-type fraction occurs and is also in line with the transition to the bicontinuous-like type of the structure in this regime. The latter represents the intermediate structural type between the TBA-dominated structure in the alcohol-rich and the water-dominated structure in the water-rich part of the system's composition.

The viscosity of the system reaches its maximum when the absolute value of the TBA-water-type HBs is at a maximum (see Fig. 6a). With a further dilution the system enters the water-continuous-like range, where the regions with TBA aggregates are locally reducing in size and are contributing less and less to the system's viscosity – the latter correspondingly decreases in this range. The TBA aggregates eventually transit to the single hydrated TBA molecules, which exhibit a relatively high self-diffusion coefficient. In parallel, all these results clearly show that even though the self-diffusion coefficients are strongly connected to the system's viscosity, it is not necessary that they fully follow its trends. Namely, the self-diffusion coefficient is a property of an individual type of molecular species that strongly depends on its specific intermolecular interactions, but the system's viscosity depends on an interplay of the effects of all the different types of molecules and their mutual interactions in the system.

### 3.4. HB Lifetime and Thermodynamics

To complement this study with the dynamic and thermodynamic aspects of hydrogen bonding, we have also analyzed the model results for the lifetime of the HBs in the TBA/water system. In Fig. 8a the probability-distribution function of the HB's lifetime for an average effective type of HB in the TBA/water system is presented, *i.e.*, all of the HBs in the system were considered at once and the average result was calculated. Up to five distinct peaks of the characteristic lifetimes can be observed in these curves, which change considerably depending on the system composition. The average HB lifetime was calculated for every curve and its resulting dependency on the composition of the system is depicted in the inset. Interestingly, the average lifetime of such an average effective HB type increases considerably when diluting the TBA with water.

In the next step we made efforts to distinguish between three individual HB types in this analysis (TBA-TBA, TBA-water, and water-water). The corresponding probability-distribution functions of the lifetimes for the individual HB types are presented in Fig. 8b-d. Some of these curves exhibit more noise than the others, because for some compositions the HBs of some type are very scarce and it was not possible to get curves with better statistics. The most interesting general conclusion is that the composition of the system has a practically negligible effect on an individual HB type lifetime, as the curves do not change much – slight changes can be inferred anyway from the average lifetime dependences in the insets. Therefore, the true reason for the considerable differences observed in Fig. 8a are different fractions of HB types in the system, as presented in Fig. 6b. With a gradual increase of the water concentration in the system the fraction of TBA-TBA HB type with the shortest characteristic lifetime gradually reduces. For more details on these probability distributions the reader is directed to the SI and Fig. S11 therein.

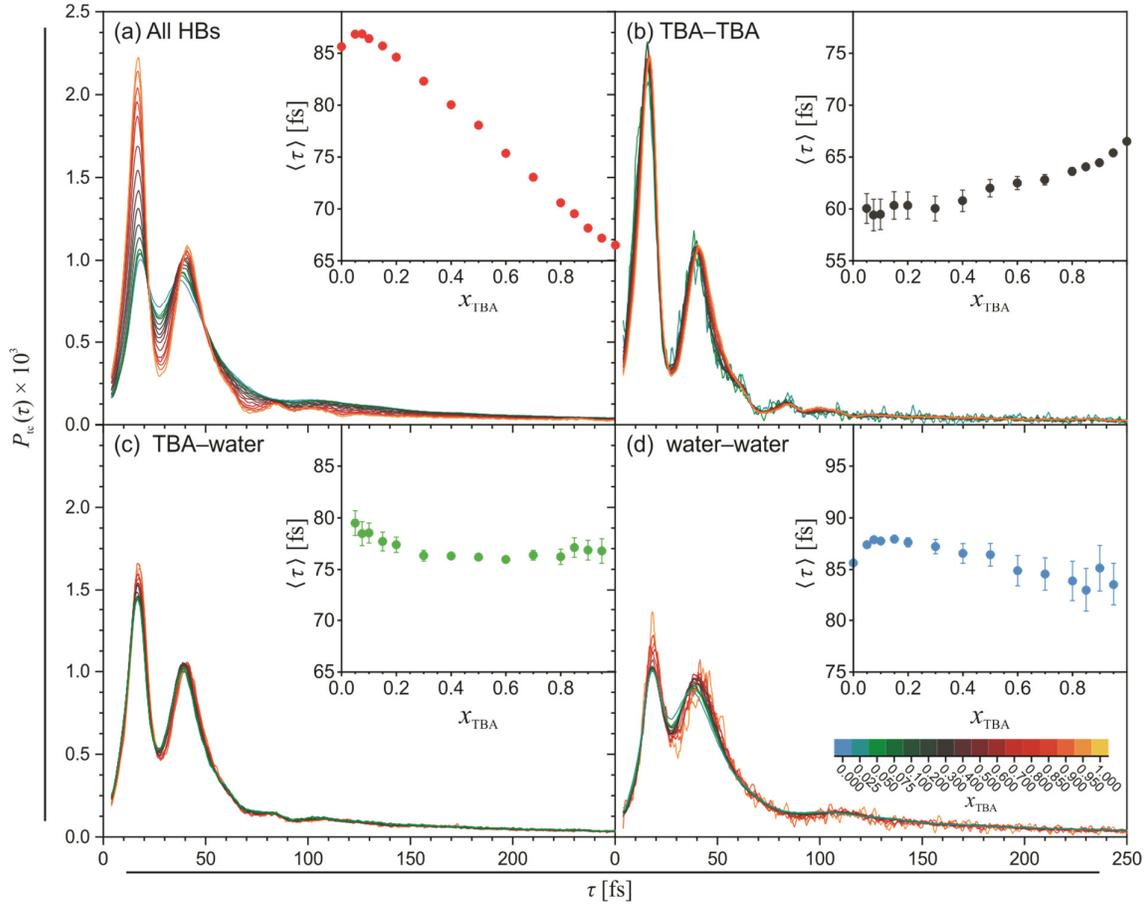

Fig. 8 Probability-distribution function of the total HB lifetime in a configuration for various system compositions for: (a) all HBs, (b) TBA–TBA, (c) TBA–water, and (d) water–water type of HBs in the system. Insets: The average HB lifetime vs. $x_{\text{TBA}}$.

The analysis of the MD-simulation results for the number of HBs depicted in Fig. S12 in the SI also provided a very interesting composition-dependent thermodynamics of the process of HB formation in the TBA/water system, which are depicted in Fig. 9. Inspecting the composition dependence of the standard Gibbs free-energy change, $\Delta G^o$, we can observe an abundant hydrogen bonding in the whole concentration range and an interesting trend of the changes of $\Delta G^o$ with the composition having a minimum (maximum absolute value) in the water-rich range around $x_{TBA} \sim 0.1$, where the formation of HBs seems thermodynamically the most favorable. Negative enthalpy changes signifying an exothermic process are almost constant in the composition range $0.1 \lesssim x_{TBA} \lesssim 0.6$ (water- and bicontinuous structural region), with the trend of increasing at lower and decreasing at higher TBA concentrations. That means that the hydrogen-bonding process is energetically somewhat more favored in the TBA-rich range than in the water-rich range. This can be reasoned with the explanation that it is energetically less favorable for the hydrophilic hydrogen-bonding groups to freely move in the more hydrophobic environment in the alcohol-rich range than in the more hydrophilic water-rich range. A very similar trend can also be observed for the entropic contribution in Fig. 9 showing that in a water-rich range the hydrogen bonding is entropically somewhat less unfavorable than in the TBA-rich range. As expected, the hydrogen-bonding is an energetically driven process and is thermodynamically slightly more favored in the water-rich range.

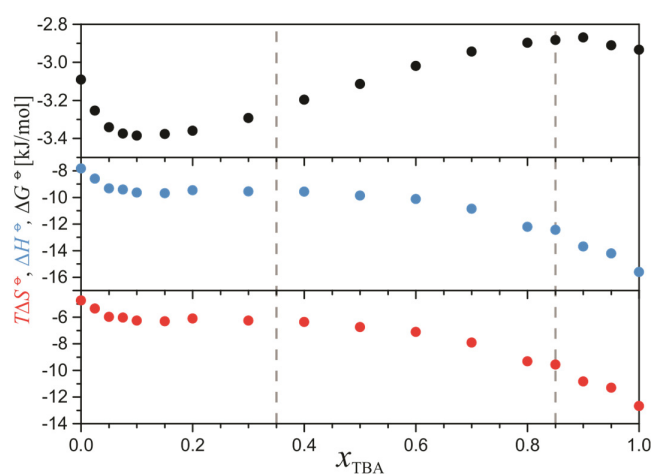

Fig. 9 The thermodynamic hydrogen-bonding properties, $\Delta G^o$, $\Delta H^o$ and $T\Delta S^o$, for the formation of the HB of an average HB type in the model TBA/water system at 25 °C.

## 4. CONCLUSIONS

The results presented in this paper illustrate that combining computer-simulation methods with the 'complemented-system approach' [28] in a structural interpretation of the experimental x-ray scattering data and in the treatment of the rheological and dynamic properties of the molecular hydrogen-bonding liquids, provides a comprehensive description of the structure-viscosity-dynamics relationship in the studied system. The latter was successfully obtained for the hydrotropic TBA in an aqueous binary system, which showed a rich variety of interesting structural and dynamic properties across the whole concentration range – four qualitatively different characteristic regimes were observed. The SWAXS and XRD techniques simultaneously revealed the structural information on the intra-, inter-, and supra-molecular scales. The MD technique helped to interpret the structural features in detail and to further relate them to the rheological data on viscosity, the time-resolved dynamic information (hydrogen-bonding lifetimes, molecular-diffusion coefficients) and the thermodynamic properties of the HB formation. With this, both important objectives of our study were met. Parts of this methodology were already previously applied in studies of molecular liquids with smaller molecules [24-33, 115, 116], but never before on a mixture of two liquids and on a system showing a very low-$q$-regime scattering-intensity increase, which are the two very important conceptual novelties of this study. Furthermore, to study the TBA/water binary system across the whole concentration range in a single study with a unified approach was a great challenge, due to the complexity of its structural and rheological behaviors – there are lots of studies on the TBA/water system available in the literature [6, 7, 44-46, 50, 57, 68, 117-120], but none with a comprehensive combination of structural, rheological and dynamics aspects and at the inter-, intra- and supra-molecular level of details, as presented here. As there is a clear need for new methodological approaches to the interpretation of SWAXS data expressed in the literature [121], this study importantly promotes the use of the complemented-system approach [28] and implicitly also other simulation-based approaches in this field [24, 25, 122-132]. Its results clearly highlight the benefits of this methodology in revealing the SWAXS details that would remain concealed without an insight into the model-based computer-simulation results. It provides an insight into the individual partial contributions to the overall x-ray scattering intensity. In this sense, the idea of 'contrast-matching', which is characteristic of the small-angle neutron-scattering domain, is extended to the x-ray scattering domain, where otherwise it cannot be applied using classic approaches [132].

One of the most interesting and even surprising structural details of the TBA/water system was revealed in this study from analyzing the results of the complemented-system approach and its simulated analogy with the contrast-matching experiment. It is connected with the fact that the initial swelling of the –OH backbone of the TBA aggregates that occurs with an increasing water concentration in the system (at $x_{\text{TBA}} \gtrsim 0.85$) does not affect the average correlation lengths between the neighboring –OH backbones, as one might expect based on the previous studies of simple aliphatic alcohols [24, 105], but rather effects the spatial orientation of the TBA molecules, which results in an increase of the average correlation lengths between the alkyl tails of the TBA molecules. In a pure TBA system the bulky *tert*-butyl groups protrude from the –OH backbone structure and orient themselves in such a way as to most efficiently occupy the space. However, with the dilution of TBA, the water molecules become incorporated into the hydroxyl network and dynamically split the –OH backbone into smaller aggregates – this gradually leads the TBA molecules to preferentially orient themselves with mutually facing hydrophobic alky tails, which in turn increases the average correlation lengths between them and can be described as the 'hydrophobic effect'. Interestingly, a very similar increase in the correlation lengths, even though for different reasons, was recently also observed for diol molecules at elevated temperatures [31]. Such a 'hydrophobic effect' is even more pronounced, when the structure transits into a bicontinuous-like type in the range $0.85 \gtrsim x_{TBA} \gtrsim 0.35$ and is progressively becoming a more and more important structural driving force (besides hydrogen bonding). Further down to $x_{TBA} \sim 0.05$, the bicontinuous structure gradually ruptures into discrete local TBA-rich regions inside the continuous aqueous medium that decrease in size and with $x_{TBA} \lesssim 0.05$ become very small and steadily transit to single hydrated TBA molecules.

We can conclude that our detailed discussion of the results revealed a very strong relationship between the hydrogen-bonding phenomena (in a static and dynamic sense), the supramolecular structural characteristics and the rheological aspects of the TBA/water system (their viscosity and molecular diffusion trends) in the whole concentration range and importantly deepened our understanding of the topic. The OPLS force field [86-88] turned out to be currently the best force field to model the TBA/water system structurally in the overall concentration range. Even though it showed some quantitative flaws in reproducing the absolute scattering intensities and the very low-$q$-regime scattering-intensity increase, it was anyway successful in a qualitative sense, predicting the latter and all the characteristic SWAXS scattering peaks. Our results have also proven that due to the presence of very large supramolecular structural features in a part of the TBA/water system's composition range ($0.3 \gtrsim x_{TBA} \gtrsim 0.05$) there is certainly the need to perform MD simulations in very large simulation boxes in order to eliminate the finite-size effects of the simulated systems.

In our future studies, we plan to exploit the recent developments in heterogeneous computing platforms with the advanced parallel-computing capabilities and challenge the presented methodology based on experimental and theoretical SWAXS data [24, 25, 28] with systems that will contain even larger structural segments of interest in the field of colloidal chemistry.

**DECLARATION OF COMPETING INTEREST**

There are no conflicts to declare.

**ACKNOWLEDGEMENTS**

We acknowledge financial support from: the Slovenian Research Agency (research core funding No. P1-0201, the project No. N1-0042 'Structure and thermodynamics of hydrogen-bonded liquids: From pure water to alcohol-water mixtures', No. BI-HU/17-18-005 'Structure of aqueous solutions of sugars and alcohols', and the project No. BI-HU/19-20-001 ' Solvation of hydroxy-containing compounds in water'), the NKFIH of Hungary (projects no. SNN116198, KH125430, TET_16-1-2016-0202, and 2018-2.1.11-TÉT-SI-2018-00016 for IP, LT and LP), the Hungarian Academy of Sciences (Janos Bolyai Research Fellowship for LT), and the Japan Synchrotron Radiation Research Institute (BL04B2, Proposal No. 2018A1132). We are most grateful to Prof. Otto Glatter for his generous contribution to the instrumentation of our Light Scattering Methods Laboratory in Ljubljana.

organization on a nanometer-scale in ternary solvent solutions containing a hydrotrope, J. Colloid Interf. Sci. 540 (2019) 623-633.

*Supporting Information for the paper:*
## Structural, Rheological and Dynamic Aspects of Hydrogen-Bonding Molecular Liquids: Aqueous Solutions of Hydrotropic *tert*-Butyl Alcohol

Jure Cerar,[a] Andrej Jamnik,[a] Ildikó Pethes,[b] László Temleitner,[b] László Pusztai,[b,c] and Matija Tomšič.[a,*]

[a.] Faculty of Chemistry and Chemical Technology, University of Ljubljana, Večna pot 113, SI-1000 Ljubljana, Slovenia.
[b.] Wigner Research Centre for Physics, Hungarian Academy of Sciences, Budapest, Konkoly Thege út 29-33., H-1121, Hungary.
[c.] International Research Organisation for Advanced Science and Technology (IROAST), Kumamoto University, 2-39-1 Kurokami, Chuo-ku, Kumamoto 860-8555, Japan.
*Correspondence e-mail: matija.tomsic@fkkt.uni-lj.si

## Appendix A – Experimental and Methods

**Chemical**

The chemical 2-methylpropan-2-ol (TBA), also known as *tert*-butyl alcohol or *t*-butanol (anhydrous, purity ≥ 99.7 %), was purchased from Sigma-Aldrich, St. Louis, Missouri, USA and used without further purification. The demineralized water was distilled in a quartz bi-distillation apparatus (Destamat Bi18E, Heraeus) and was then used to prepare aqueous TBA solutions (specific conductance of the water was less than $6.0 \cdot 10^{-7}$ $\Omega^{-1}$cm$^{-1}$) that were analyzed at 25 °C. Throughout the text the composition of the aqueous TBA samples is presented as a molar fraction of TBA, $x_{\text{TBA}}$.

**Small- and Wide-Angle X-Ray Scattering**

The SWAXS measurements were performed on a modified Kratky camera (Anton Paar KG, Graz, Austria) with a conventional X-ray generator (GE Inspection Technologies, SEIFERT ISO-DEBYEFLEX 3003). The incident beam with the wavelength $\lambda = 0.154$ nm was generated utilizing a Cu anode operating at 40 kV and 50 mA. It was monochromatized and focused on a Göbel mirror and passed the block-collimation system to result in a line-collimated monochromatic primary beam. The samples were placed in a standard quartz capillary (outer diameter of 1 mm and wall thickness of 10 μm) and thermostated at 25 °C using a Peltier element. One-hour measurements were performed utilizing a 2D-imaging plate system (Fuji BAS 1800II) with a spatial resolution of 50×50 μm$^2$ per pixel and a scattering vector, $q$, ranging from 0.1 to 25.0 nm$^{-1}$. The scattering vector is defined as $q = 4\pi/\lambda \sin(\vartheta/2)$, where $\vartheta$ is the scattering angle. The obtained data were corrected for X-ray absorption and capillary scattering and were put on an absolute intensity scale using water as a secondary standard [1]. The resulting data were still experimentally smeared due to the finite dimensions of the primary beam [2].

**Synchrotron X-Ray Diffraction**

The BL04B2 high-energy X-ray diffractometer [3], installed at the SPring-8 synchrotron facility (Hyogo, Japan), was used to perform wide-angle x-ray diffraction measurements on some of the TBA/water mixtures. X-ray photons with 61.3 keV energy, corresponding to a wavelength of 0.02023 nm, made a wide range of the scattering vector available: $1.6 \lesssim q \lesssim 200$ nm$^{-1}$. Thin-walled quartz capillaries with an outer diameter of 2 mm were used to contain the liquid samples during the XRD measurements. Raw experimental XRD data were treated according to the standard procedures [4], *i.e.*, they were normalized by the incoming primary beam intensity, corrected for the absorption, polarization and contributions from the empty capillary. Patterns that span the entire $q$-range were obtained by normalizing and merging each frame in electron units and removing the inelastic (Compton) scattering contributions.

**Molecular Dynamics Simulations**

MD simulations were performed using the GROMACS software package (version 2018.2) [5-7] and applying the OPLS-AA *tert*-butyl model [8-10] and TIP4P or TIP4P/2005 water models [11, 12]. This TBA model showed the best structural performance in the case of a few test compositions of the TBA/water system spread over the whole composition range, *i.e.*, it yielded the best agreement between the calculated and the experimental SWAXS curves in terms of matching the scattering peak positions; the TraPPE [13], CHARMM [14], and GROMOS [15] force fields were also tested. MD simulations were conducted in the canonical ensemble (NVT) in a cubic simulation box with periodic boundary conditions. The number of molecules in a simulation box was set to match the experimental densities from the literature [16]. The simulation boxes of two different sizes were utilized: (*i*) box with a side length of ~30 nm for the compositions ranging $0.05 < x_{\text{TBA}} < 0.3$ and (*ii*) box with a side length of ~10 nm. The larger boxes were needed to ensure that all the pair-correlations faded over the distances comparable to the half-length of the simulation box – this half-length corresponds to the highest dimension up to which the structural information is still reliable. The equations of motion were integrated using the Verlet leapfrog algorithm with a time step of 2 fs [17]. The short-range interactions were calculated using the Verlet neighbor scheme with a single cut-off distance of 1.4 nm and long-range dispersion corrections for the energy and pressure. The long-range Coulombic interactions were handled by the Particle mesh Ewald method with cubic interpolation and a grid spacing of 0.12 nm. The temperature was controlled using a Noose-Hoover thermostat at 298.15 K with a relaxation time of 3.0 ps [18]. The hydrogen covalent bond lengths were constrained using the LINCS algorithm [19].

The initial configuration of the simulation box was constructed utilizing Packmol software (version 17.221) [20, 21]. The minimization procedure was performed using a 1000-step steepest-descent algorithm [22]. The simulated systems were equilibrated by a subsequent MD run corresponding to the simulated time of 5.0 ns and followed by the data-collection run – a configuration was saved in a time-step corresponding to 25 ps, which yielded 100 independent simulation-box configurations used in the subsequent calculations of the SWAXS intensities and the statistical analysis of the MD simulation results. All the presented snapshots of the simulation-box configurations and molecules were rendered using VMD software [23, 24].

From the MD simulation results the self-diffusion coefficients, $D_i$, for each i-th molecule were obtained utilizing the relation [25]:

$$\lim_{t \to \infty} \left\langle \left| \vec{r}_i(t) - \vec{r}_i(0) \right|^2 \right\rangle = b_o + 6D_i \cdot t, \quad (S1)$$

where $\vec{r}_i(0)$ and $\vec{r}_i(t)$ are the initial position and the position at time t of the i-th particle in the system, respectively, and $b_o$ is the regression parameter. They were further used to access the viscosity of the system according to the following semi-empirical expression [26-28]:

$$\eta \approx RT \cdot \sum_{i=1}^{N_T} \xi_i c_i \overline{V}_i^{-\frac{2}{3}} \cdot \frac{1}{D_i}, \quad (S2)$$

where $\eta$ is the viscosity, $R$ is the gas constant, $T$ is the temperature, $c_i$, $V_i$ and $D_i$ are the molar concentration, partial molar volume and the self-diffusion coefficient of the i-th component, respectively. The parameters $\xi_i$ are the constants that are to be determined to make the equation exact for the pure solvent and for solutions that contain the maximum concentration of each solute in pure solvent.

To calculate the model-based viscosity of the systems the non-equilibrium MD Periodic Perturbation Method (PPM) was employed – the method is explained in detail in the literature [29, 30]. For the PPM simulations the same basic settings were used as for other MD simulations. A periodic cosine acceleration profile with acceleration amplitudes, $\Lambda$, ranging from 0.005 to 0.1 nm·ps$^{-2}$ was applied to pre-equilibrated NVT configurations. The system was equilibrated for a time corresponding to 5.0 ns to allow the velocity profile to fully develop, followed by a subsequent data-collection period representing 7.5 ns of the system's time development. The non-equilibrium MD simulations were also performed using the GROMACS molecular-simulations package. To relate the calculated viscosity values to the experimentally available ones the zero-shear viscosity is extrapolated using the Carreau-Yasuda model [31]:

$$\eta(\dot{\gamma}) = \frac{\eta_o - \eta_\infty}{\left[1 + (\lambda \dot{\gamma})^a\right]^{\frac{1-n}{a}}} + \eta_\infty, \quad (S3)$$

where $\dot{\gamma}$ is the shear rate, $\eta_o$ is the Newtonian or zero-shear viscosity, $\eta_\infty$ is the infinite-shear viscosity, $\lambda$ is a time constant, $n$ is the power-law exponent that describes the viscosity in the shear thinning regime, and $a$ is a dimensionless parameter that describes the transition region between the zero-shear and the power-law region.

### Radial Distribution Functions and Spatial Distributions

The radial distribution functions (RDFs), $g_{ij}(r)$ give complete information about the spatial arrangement of the molecular pairs i–j only for the spherically symmetrical interparticle potential of interaction [32], *i.e.*, it does not reveal information about the mutual interparticle orientation. As we are investigating spherically non-symmetric molecular systems, where the mutual orientation of different parts of the molecules affect the supramolecular self-assembly structure, we also show spatial distributions (SDs) to depict the intermolecular spatial organization in 3D. For this we need to ensure that all of the molecules considered in the analysis of an individual SD are in the same frame of reference. Therefore, as explained in details in ref. [33], we need three reference points (the central atom, $C$, the first reference atom, $R_1$, and the second reference atom, $R_2$) to define the new coordinate system using the orthonormal basis: $\vec{i} = \overrightarrow{CR_1}/|\overrightarrow{CR_1}|$, $\vec{k} = \overrightarrow{CR_2} \times \overrightarrow{CR_1}/|\overrightarrow{CR_2} \times \overrightarrow{CR_1}|$, and $\vec{j} = \vec{k} \times \vec{i}$, where $\vec{i}, \vec{j}$, and $\vec{k}$ are the unit vectors pointing in the directions of the $x$, $y$, and $z$ axes, respectively (see Fig. 2 in ref. [33]). A color scale is assigned to the spatial point according to its occurrence probability value (only points with the value greater than 1 are depicted) and some cut-off distance is normally used for the sake of clarity, *i.e.*, when presenting the SD related to some individual peak in the RDF the two cut-off distances are marked with vertical lines (*e.g.*, see Fig. 3a in the paper). The SDs in the video clips presented in this SI were rendered using ParaView software [34].

### Calculating the partial contributions to the SWAXS intensity – theoretical analogy with 'contrast matching'

Omitting a certain type of atoms in the resulting MD simulation box and applying Eq. (1) from the paper, we can calculate the different partial scattering contributions to the total SWAXS curve. The partial contributions $I_{\text{TBA}}(q)$ and $I_{\text{H}_2\text{O}}(q)$, depicted in Fig. 2a and 2b, were obtained by considering only the atoms of TBA and the atoms of water in the simulation box, respectively. Similarly, the cross-term contribution between the molecules (or parts of the molecule) of the type X and Y, $H_{\text{X-Y}}(q)$, was obtained by taking into account only those double-sum terms that contain atom i from the molecule of type X and atom j from the molecule of type Y, and omitting the second term in Eq. (1) from the paper, yielding the equation:

$$H_{\text{X-Y}}(q) = \frac{2}{V} \left\{ \sum_{\substack{i=1 \\ i \in X}}^{N} \sum_{\substack{j=1 \\ j \in Y}}^{N} b_i(q) b_j(q) \cdot \frac{\sin(q r_{ij})}{q r_{ij}} \cdot H(r_c - r_{ij}) \right\}, \quad (S4)$$

Correspondingly, the following relations hold for the terms calculated and presented in Fig. 2 in the paper and Fig. S5 later in the text:

$$\begin{aligned} I_{\text{total}}(q) &= I_{\text{TBA}}(q) + I_{\text{H}_2\text{O}}(q) + H_{\text{TBA-H}_2\text{O}}(q) \\ I_{\text{TBA}}(q) &= I_{-\text{C(CH}_3)_3}(q) + I_{-\text{OH}}(q) + H_{\text{OH-C(CH}_3)_3}(q) \end{aligned}, \quad (S5)$$

where $I_{total}(q)$ is the total SWAXS intensity calculated by Eq.e (1) from the paper, and $I_X(q)$ is the partial scattering contribution calculated with Eq. (1) from the paper considering only the molecules (or molecular part) of type X in the simulation box.

**Hydrogen-Bond Analysis**

The three geometrical criteria for consideration of the hydrogen-bond (HB) are schematically depicted in Fig. S1 and are the following: (*i*) the distance between the hydrogen-bonded hydrogen and oxygen, $R_{OH} \leq 0.26$ nm (*ii*) the distance between the two hydroxyl oxygen, $R_{OO} \leq 0.35$ nm and (*iii*) the angle between the oxygen, the covalently bonded hydrogen, and the neighboring oxygen atoms, $\phi_{OOH} \leq 30°$ [33, 35, 36]. The values were determined from the positions of the minimum after the first maximum in the corresponding radial distribution functions of aqueous TBA with $x_{TBA} = 0.5$ (see Fig. 3b in the paper) and did not change noticeably with $x_{TBA}$. The time-dependent histogram of HB presence (positive values) and absence (negative values) depicting the spurious formation and spurious breaking of the HB on the time scale of hydrogen-bond lifetime $\tau$ is also depicted in the bottom of Fig. S1.

Analyzing the lifetimes of the three different HB types present in the system, we followed the work of Voloshin and Naberukhin [37] calculating the probability distribution function of the total HB lifetimes in a configuration, $P_{tc}(\tau)$, *i.e.*, the fraction of HBs that have a total lifetime $\tau$ in the statistically representative configuration. The spurious breaking and spurious formation of the HBs were also considered in a way that the breaking or formation of the HB for less than 4 fs was not counted as a valid event. The average lifetime for a certain type of HB was calculated according to the equation:

$$\langle \tau \rangle = \frac{\sum_i P_{tc}(\tau_i) \cdot \tau_i}{\sum_i P_{tc}(\tau_i)}, \quad (S6)$$

where lifetimes of up to 500 fs were still considered, as the function $P_{tc}(\tau)$ did not fully decay earlier.

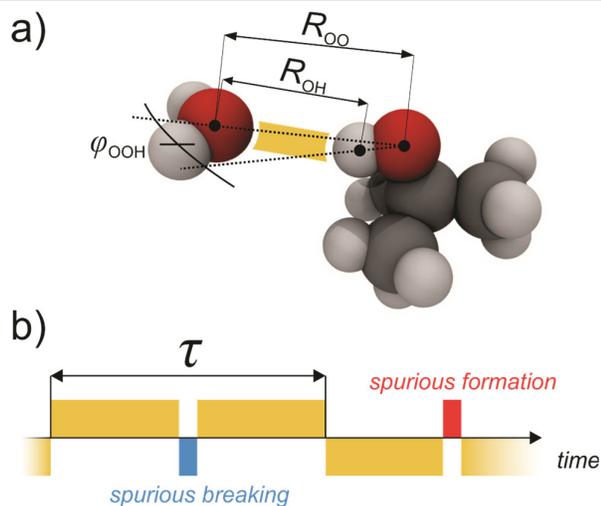

Fig. S1. Schematic representation of the criteria for consideration of presence and lifetime of the hydrogen bond.

The thermodynamic quantities for the formation of an average HB per particle in the simulated system were evaluated in analogy with the approach described by Van der Spoel *et al.* [38]. It is based on calculating the average number of HBs per molecule, $N_{HB}$, and the maximum number of HBs per molecule, $N_{max}$, whose values determine the equilibrium constant for the formation of HB, $K_{eq}$:

$$K_{eq} = \frac{N_{HB}}{N_{max} - N_{HB}}, \quad (S7)$$

and further on using the standard thermodynamic relations for the changes of the standard Gibbs free energy, $\Delta G^\ominus$, enthalpy, $\Delta H^\ominus$, and entropy, $T\Delta S^\ominus$, for the process [38]:

$$\Delta G^\ominus = k_B T \cdot \ln K_{eq}$$
$$\Delta H^\ominus = \frac{\partial (\Delta G^\ominus / T)}{\partial (1/T)} \quad . \quad (S8)$$
$$T\Delta S^\ominus = \Delta H^\ominus - \Delta G^\ominus$$

In strict thermodynamic language, the equilibrium constant in Eq. (S7) should be expressed with activities of reactants and products. These activity values depend on the activity coefficients and the choice of the standards states. However, $K_{eq}$ in Eq. (S7) provides information about the average fraction of HB donors participating in the HB formation, and should be considered as an apparent equilibrium constant.

The values for $N_{HB}$ were obtained from the resulting configurations in the MD simulation boxes. The theoretical values for $N_{max}$ of the individual TBA-TBA ($N_{max}^{(TBA-TBA)}$), water-water ($N_{max}^{(w-w)}$), mixed TBA and water ($N_{max}^{(mix)}$), and average HB type ($N_{max}^{(average)}$) were calculated on the basis of the probabilities of the formation of HB in an ideal solution with a uniform distribution of molecules under the assumption that the TBA molecule can donate only 1 HB and the water molecule up to 2 HBs. The expressions read [38]:

$$N_{max}^{(TBA-TBA)} = \frac{N_{TBA} - 1}{N_{TBA} - 1 + 2N_w} \quad (S9a)$$

$$N_{max}^{(w-w)} = \frac{2(2N_w - 2)}{N_{TBA} + 2(N_w - 1)}$$
$$N_{max}^{(mix)} = \frac{2N_{TBA} N_w (2N_{TBA} + 4N_w - 3)}{(N_{TBA} + N_w)(N_{TBA} - 1 + 2N_w)(N_{TBA} + 2N_w - 2)}, \quad (S9b)$$
$$N_{max}^{(average)} = \frac{N_{TBA} + 2N_w}{N_{TBA} + N_w}$$

and were derived using the probabilities that a selected type of molecule, TBA or water, which is first taken from the system and then inserted into the system again forms a bond with the other type of molecule, again TBA or water. These probabilities thus do not correspond to the probabilities of whether the bonds are formed at all, but under the assumption that the bonds are certainly formed, only address the question as to the chances for the bonds with different types of molecules. For instance, when we select a TBA molecule with one possible bond then after inserting it into the system we only consider the chances that it forms one bond with either another TBA molecule or a water molecule. Thus, the probability of the formation of a TBA-TBA bond is equal to the ratio between the number of possible bonds of the selected TBA molecule with the remaining TBA molecules – this number is $(N_{TBA} - 1)$ – and the number of all the possible bonds, where we have to take into account also $2N_w$ possible bonds with $N_w$ water molecules.

The maximum number of HBs of the mixed TBA and water type is obtained from the probabilities of the TBA molecule to HB with a water molecule, $N_{max}^{(TBA-w)}$, and vice versa, $N_{max}^{(w-TBA)}$:

$$N_{max}^{(TBA-w)} = \frac{2N_w}{N_{TBA} - 1 + 2N_w}$$

$$N_{max}^{(w-TBA)} = 2 \cdot \frac{N_{TBA}}{N_{TBA} + 2(N_w - 1)} \quad , \quad (S10)$$

weighted with the molar fractions of TBA and water, $x_{TBA}$ and $x_w$, respectively:

$$N_{max}^{(mix)} = x_{TBA} \cdot N_{max}^{(TBA-w)} + x_w \cdot N_{max}^{(w-TBA)} . \quad (S11)$$

The average maximum number of HBs per particle is then equal:

$$N_{max}^{(average)} = x_{TBA} \cdot N_{max}^{(TBA-TBA)} + x_w N_{max}^{(w-w)} + N_{max}^{(mix)} . \quad (S12)$$

## Appendix B – Experimental and Methods

### Experimental and Calculated SWAXS Results

The calculated curves presented in Fig. 1 are based on the OPLS-AA force-field TBA model [8-10] and the TIP4P/2005 force-field water model [12]. The calculated scattering curves presented in Fig. 1b are numerically smeared and are as such directly comparable to the experimental data. In general, they follow from the MD results for the simulation boxes with a side length of 10 nm, but to provide a somewhat better resolution in terms of objects of larger dimensions for the concentration range with $0.3 \gtrsim x_{TBA} \gtrsim 0.05$, much bigger simulation boxes with a side length of 30 nm were used. The results for two of the curves from this concentration range are also depicted in Fig. S2 on a normal scale.

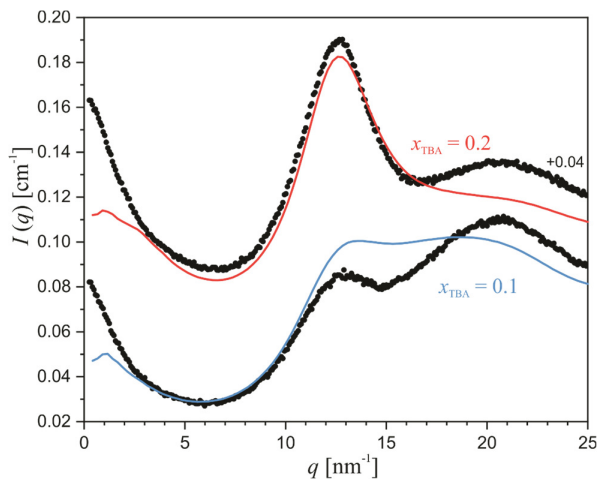

Fig. S2. The experimental (symbols) and calculated (lines) SWAXS intensities at $x_{TBA} = 0.1$ (blue) and 0.2 (red) according to the OPLS-AA [8-10] and TIP4P/2005 [12] force fields.

Unfortunately, the intensity increase in the calculated curves was not sufficient and even a shallow maximum appeared just below the sharp continuous intensity increase of the experimental curve in case of $x_{TBA} = 0.1$ and 0.2. Therefore, we tested the united-atom TraPPE force-field in these cases [13], as it was the most promising one to better reproduce this intensity increase and depict the results in Fig. S3. However, it turned out that in contrast to the MD simulation results in a 10-nm simulation box, which are represented in Fig. S3a, the TraPPE model system simulated in a 30-nm box was not stable – the micro phase segregation was steadily increasing during the simulation and the system was driven towards the macro phase-separation, as evident from the snapshots during the time-demanding MD simulation depicted in Fig. S3b. Nevertheless, these results show that the very low-$q$-regime scattering intensity increase originates in the supramolecular level of the structure and in parallel also strongly supports the findings of Gupta and Patey [39] on the necessity of the simulations in large simulation boxes to ensure better sampling of the model system.

Eventually, we decided to base our study on the results of the OPLS-AA force-field *tert*-butyl model throughout the whole concentration range, as presented in Fig. 1 in the paper. Although it somewhat underestimated the absolute scattering intensities, it generally yielded the best overall agreement in the scattering peak positions, as is evident also in Fig. S4. Besides the latest TIP4P/2005 force field, also the older version TIP4P was tested and yielded very similar structural results (not shown).

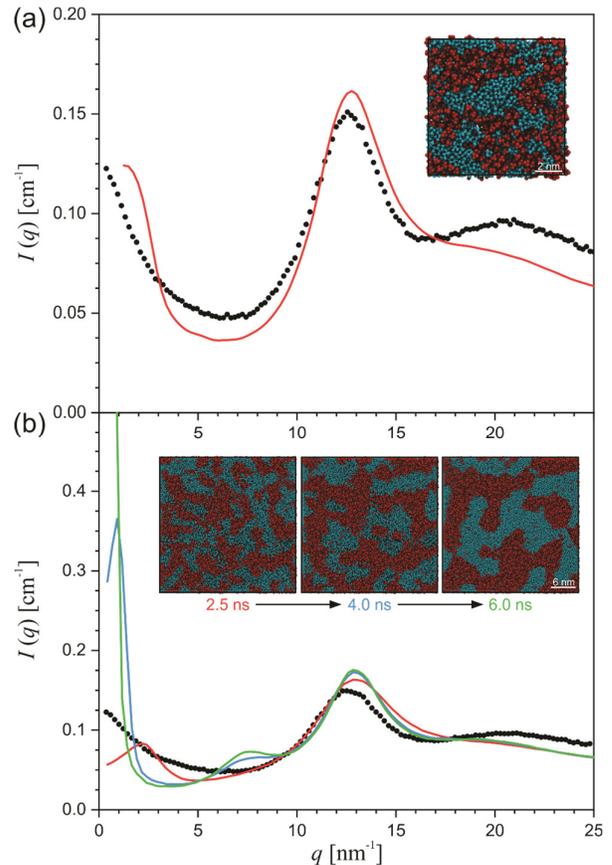

Fig. S3. The experimental (symbols) and the calculated (line) UA-TraPPE and TIP4P/2005 force-field-based SWAXS intensities for the TBA/water system at $x_{TBA} = 0.2$. The results for simulations in a simulation box with the side length of (a) 10 nm and (b) 30 nm are shown.

We can claim that a good qualitative agreement between the calculated and experimental data can be observed in Fig. 1a in the paper, where the SWAXS data also present the range of very low values of the scattering vector $q$ in more details. We must point out that the presence of the scattering peaks, especially their relative positions, and also the overall trends in the course of the SWAXS curves, are in the first instance of much higher importance for the structural interpretations than the absolute values of the SWAXS intensities. The former reflect the structural correlation lengths and sizes of the scattering moieties, whereas the latter rather strongly depend on the values of the scattering contrast, which is hard to reproduce correctly by the physical models that are always simplified to some level [33, 35, 40]. Accordingly, the calculated SWAXS curves successfully predict the scattering-intensity increase in the very low-$q$-regime at low TBA concentrations, as also seen for $x_{TBA} = 0.1$ and $0.2$ somewhat better in Fig. S2, although it seems that such an increase is not sufficient in the absolute sense in comparison to the experimental data. Nevertheless, we can conclude that the applied OPLS-AA and TIP4P/2005 force-field models also successfully reproduce the general structural characteristics of the studied TBA/water system on the inter- and supra-molecular level.

**Partial SWAXS Contributions**

The detailed partial scattering contributions to the partial SWAXS intensity contribution of the TBA molecules are presented in Fig. S5.

The 3D version of spatial distributions, which lead to the (*i*) and (*ii*) peaks in Fig. 3a in the paper for the system with $x_{TBA} = 0.5$ are presented in the video clips Video S1 and S2, respectively. In video clip Video S3 the concentration dependence of the (total) spatial distribution of tertiary C atom around the central TBA molecule is presented. We can see that with decreasing $x_{TBA}$ the 'clouds' that are close to –OH group weaken, as the clouds close to the alkyl part of the TBA molecules gain on the size and intensity. This clearly indicates on the decreasing tendency of the TBA-TBA type HB formation and the increasing hydrophobic effect with increasing water concentration in the system, *i.e.* on the competition between hydrogen bonding between the TBA molecules and the hydrophobic effect in governing the structure of aqueous TBA.

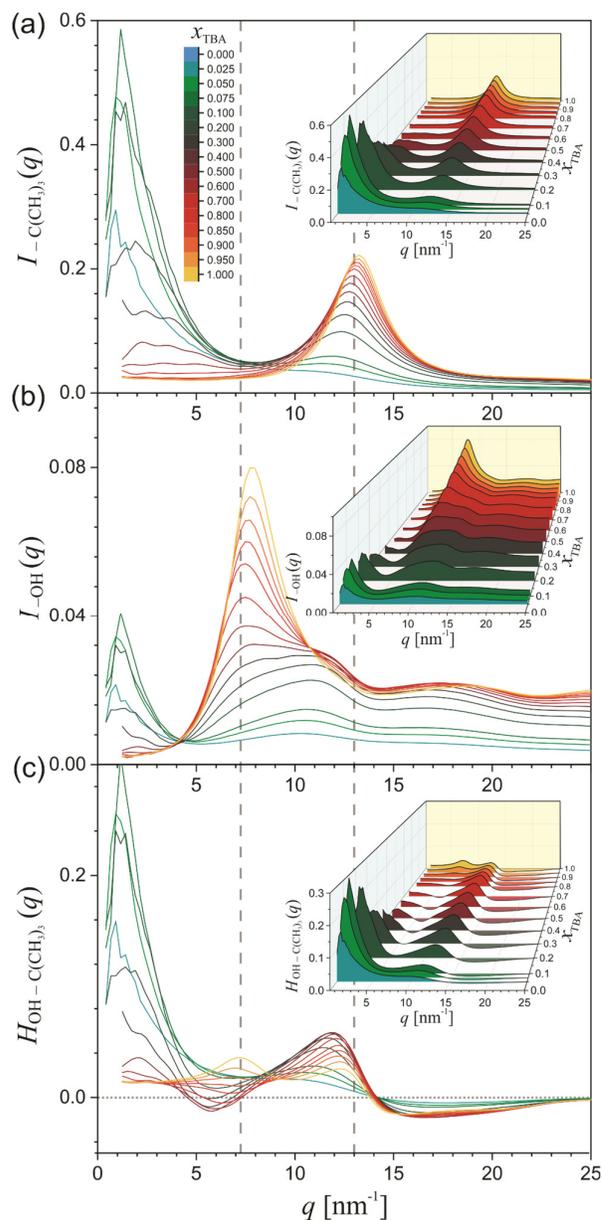

Fig. S5. The detailed partial scattering contributions to the theoretical SWAXS intensities of the TBA/water system at various compositions: (a) $I_{-C(CH_3)_3}(q)$, (b) $I_{-OH}(q)$, and (c) $H_{OH-C(CH_3)_3}(q)$. The vertical lines are drawn at 7.4 and 13 nm$^{-1}$.

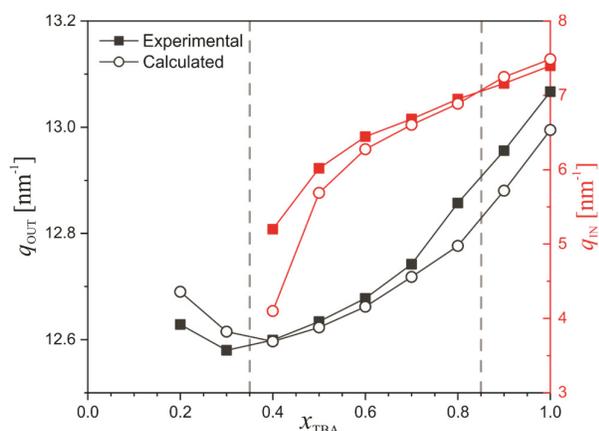

Fig. S4. The dependence of the inner (red), $q_{IN}$, and outer (black), $q_{OUT}$, scattering peak position on the system's composition obtained from the experimental (squares) and calculated (circles) SWAXS curves from Fig. 1a. The vertical lines are drawn at 0.35 and 0.85.

In Fig. S6 the distribution of the dihedral angle H-O-C-C in the model TBA molecule is presented and shows that there are no significant intra-molecular changes observed for TBA with changes to the water concentration in the system.

SUPPORTING INFORMATION

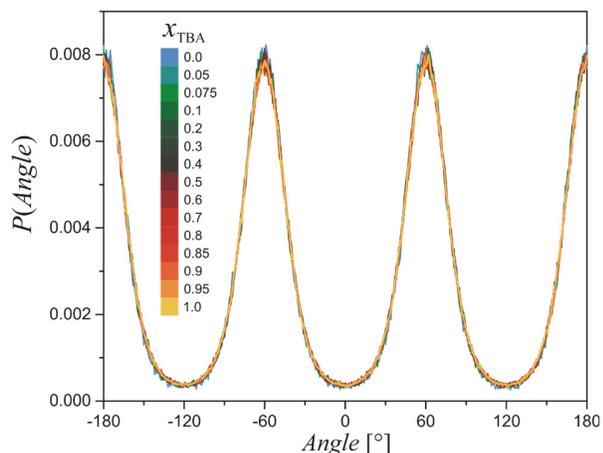

Fig. S6. The distribution function of the dihedral angle H-O-C-C in the modelled TBA/water system at various compositions.

**Low-$Q$ Increase of the SWAXS Intensity**

Investigating the sharp scattering intensity increase in the low $q$-regime at compositions $0.3 \gtrsim x_{TBA} \gtrsim 0.05$ in Fig. 1 and at compositions $x_{TBA} \lesssim 0.85$, the partial scattering contribution in Fig. 2a and 2b further, we assessed the local scattering contrasts within the TBA-rich and water-rich regions of the mixture across the whole composition range of the system. We used the experimental density data for the TBA/water mixture [41] and calculated the average electron densities for different system's compositions. Similarly, we used the experimentally determined partial molar volume data for the TBA/water mixture [41] to calculate the corresponding local electron densities within the TBA-rich and water-rich regions for different system's compositions – the assumption that we needed to make was that such localized homogeneous regions form in the system at all compositions. The results are presented in Fig. S7.

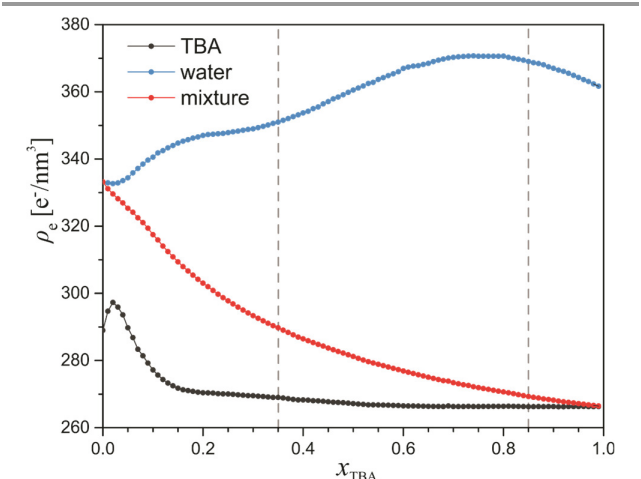

Fig. S7. The calculated electron density, $\rho_e$, i.e., the local scattering contrast in TBA- (black) and water-rich regions (blue) in the TBA/water system at various compositions. The average electron-density profile (red) is also shown. The experimental values of the densities and the partial molar volumes, necessary for the calculations of these electron density profiles, were taken from the literature [41]. The vertical lines are drawn at 0.35 and 0.85.

The difference between the local and average electron density determines the local scattering contrast, which is of utmost importance to ensure that the structural segments are at all seen by the SWAXS method (only the scattering segments with sufficient scattering contrast affect the interference pattern). The other two properties that are also important in this sense are the size and the concentration of the scattering structural segments. It is interesting to notice in Fig. S7 that already at very low $x_{TBA}$ values the two local scattering contrasts are considerable and that they become equal (but opposite in sign) already at $x_{TBA} \approx 0.15$. In connection with this, we must remind the reader that the well-known Babinet's principle states that the scattering pattern of a homogeneous body with positive scattering contrast in a homogeneous medium is identical to that of a corresponding hole (pore) in a homogeneous medium with the appropriate negative contrast [42, 43]. In this sense, we can understand the sharp positive increase in both the $I_{TBA}(q)$ and $I_{H_2O}(q)$ contributions at very low $q$-values as also the negative $H_{TBA-H_2O}(q)$. Furthermore, as this intensity increase emanates from the presence of the TBA molecules in these water-rich samples and their influence on the structure of the system, the fact that it is (surprisingly) even more pronounced in absolute values in the $I_{H_2O}(q)$ than in the $I_{TBA}(q)$ contributions, can only be the consequence of the Babinet principle. Namely, calculating the $I_{H_2O}(q)$ contribution only the water molecules are present in the simulation box of these water-rich systems, which causes the presence of the "holes" with zero electron density in the regions where the TBA molecules reside otherwise. Correspondingly, in addition to the general contribution of the water structure to the absolute values of $I_{H_2O}(q)$ in Fig. 2b, also these holes with great scattering contrast contribute strongly.

The size and concentration of the scattering structural segments, which can be inferred from the schemes in Fig. 4, is also very important. Therefore, it is worth mentioning that at $x_{TBA} \approx 0.2$, where the sharp intensity increase is the most pronounced in the overall SWAXS curve, the weight fraction of TBA in the system is already 0.5 and the local TBA-rich regions already reach an almost comparable effective size and concentration as the local water-rich regions (see Fig. 4c). With a further increase of $x_{TBA}$ the TBA molecular organization gradually begins to dominate the system's structure – in parallel the partial scattering contributions practically mutually cancel at very low-$q$-values (the transition from the water-dominated structure to the TBA-dominated structure occurs over a bicontinuous-like type of the micro-phase-separated structure – see Fig. 4b).

**TBA Aggregates**

In Fig. S8 the distribution functions of the TBA aggregate size for the TBA/water system throughout the whole concentration range are shown. Interestingly, in the TBA-rich range a maximum could be observed at the aggregate size of 4 TBA molecules. We could see that this maximum appears because of the larger fraction of cyclic aggregates at higher TBA concentrations in the system – the 4-membered –OH cyclic aggregates were found to be the most abundant and were reducing in concentration with increasing water concentration in the system.

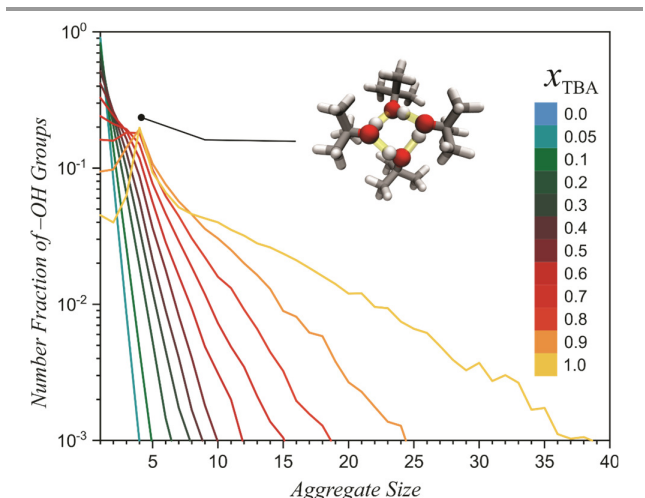

Fig. S8. The TBA hydroxyl aggregate size distribution in the TBA/water system at different compositions. The water molecules were not considered to be a part of these aggregates.

**Hydrogen-Bonding in the TBA/Water System**

In Fig. S9 we show the concentration dependence of the average number of HBs per molecule in the model TBA/water system, considering both donor and acceptor HBs per molecule. It is worth pointing out that depending on the TBA/water system's composition, the water molecules have in total from approximately 1.1 to 1.3 more HBs per molecule than the TBA molecules in the TBA/water system. Furthermore, it is interesting to observe that the deviations from the linear trend of these curves appear around $x_{TBA} \sim 0.35$ and $x_{TBA} \sim 0.85$ in Fig. S9, which are again the characteristic limiting concentrations also observed for different structural regimes.

**Viscosity and Self-Diffusion Coefficients**

In Fig. S10 the viscosity curves assessed by the semi-empirical Eq. (S2) and the calculated model-based self-diffusion coefficient values are depicted and compared with the available experimental data [16]. As the parameters $\xi_i$ in Eq. (S2) are determined in a way that the equation is exact for pure water and TBA in our case,

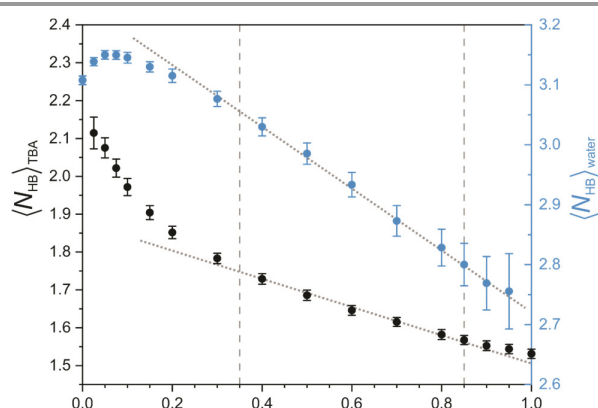

Fig. S9. The dependence of the average number of HBs per molecule, $\langle N_{HB} \rangle$, for TBA (black) and water (blue) molecules on the system's composition. All the different types of HB per molecule were considered in this analysis. The vertical lines are drawn at 0.35 and 0.85.

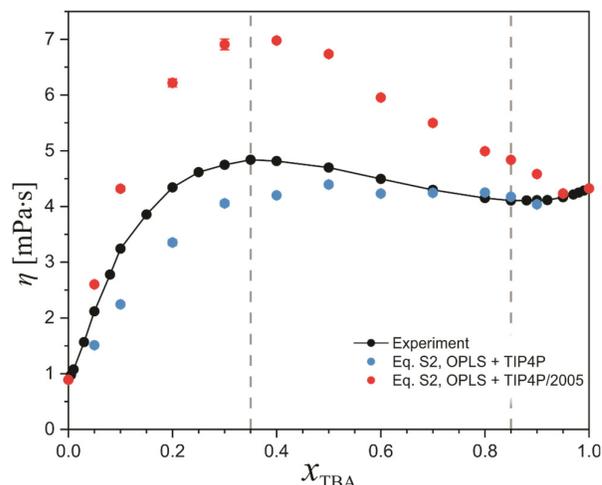

Fig. S10. The calculated viscosity values assessed by Eq. (S2) and the calculated model-based self-diffusion coefficient values and the available experimental data [16]. The vertical lines are drawn at 0.35 and 0.85

these two experimental viscosity points coincide with the calculated ones – the viscosity values in-between depend on the calculated model self-diffusion coefficient values. As we can see, the trend of viscosity changes is also qualitatively very well reproduced on the basis of these model-based self-diffusion coefficient values.

**Hydrogen Bond Lifetime and Thermodynamics**

In Fig. S11 we show the probability-distribution function of the total HB lifetime for TBA–TBA type HBs in pure TBA, water–water type HBs in pure water and TBA–water type HBs in TBA/water mixture with $x_{TBA} = 0.5$. In general, we can observe as many as five characteristic peaks for the TBA-TBA type of HB (at 16.4, 41.1, 57.6, 82.0 and 102.5 fs) in Fig. S11a and four characteristic peaks for the water-water type (at 17.5, 36.4, 55.6 fs and another very shallow and broad one at 98.6 fs) in Fig. S11b. Similar situation with five characteristic peaks is observed also in Fig. S11c for the TBA–water type HBs (at 16.5, 40.0, 56.7, 79.2 and 106.9 fs).

Each such peak represents a different population of HBs with a characteristic HB lifetime. One would expect that each reflects some specific characteristic molecular hydrogen-bonding situation or environment. However, putting in some effort to analyze the MD trajectory we were unfortunately not successful in recognizing and connecting these specific HB lifetimes with some specific local structural features in the system. Therefore, this remains to be answered in some specialized future study. We should also point out that such results are expected to depend on the model used and on the criteria for considering the HB, of course.

We now turn to the thermodynamics of hydrogen-bonding. In Fig. S12 we show the difference between the number of HBs per molecule in the model system obtained from our MD simulations and its theoretical maximum values calculated by Eq. (S9). The fact that in a part of the composition range some of these quantities adopt the positive values (surplus of the HBs), indicates that the maximum hydrogen-bonding values are not accessed correctly for these cases – the reason lies in the fact that the simulated (and also real) system is certainly not statistically random, but rather locally hydrophilic and hydrophobic environments appear in the system on a microscopic level. Therefore, it is unfortunately not possible to reasonably treat the thermodynamic properties of an individual type of HB, but rather only the properties of the average HB type – the later can even be studied experimentally [38].

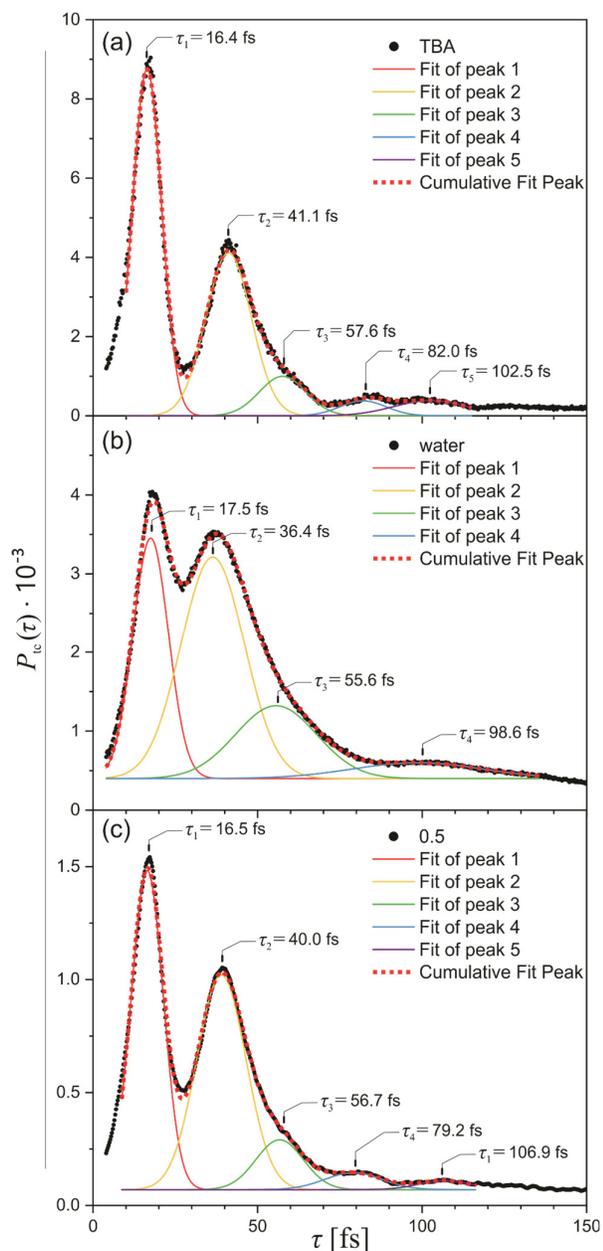

Fig. S11. Probability-distribution function of the total HB lifetime in a configuration, $P_{tc}(\tau)$, for (a) TBA–TBA type HBs in pure TBA, (b) water–water type HBs in pure water and (c) TBA–water type HBs in TBA/water mixture with $x_{TBA} = 0.5$, disassembled to a number of Gaussian distributions, each corresponding to an individual population of HBs with a characteristic average lifetime given next to the distribution.

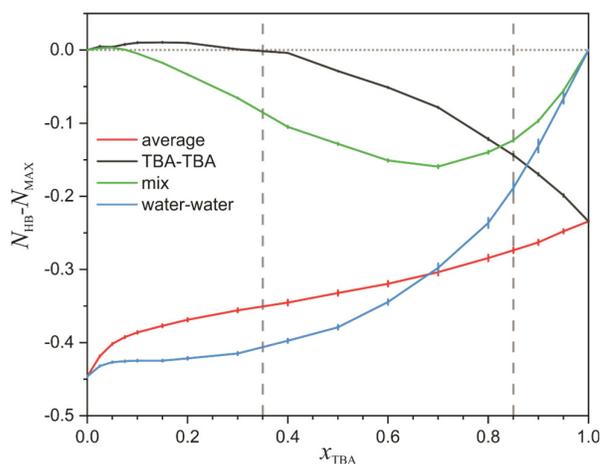

Fig. S12. The difference between the number of HBs per molecule in the model system and its statistical maximum for different HB types: TBA-TBA (black), mixed TBA and water (green), water-water (blue); and for average HB in the system (black). The vertical lines are drawn at 0.35 and 0.85.